\documentclass[aps,twocolumn,superscriptaddress,10pt,floatfix]{revtex4-2}
\usepackage{braket}
\usepackage{comment}
\usepackage{graphicx}
\usepackage{soul}
\usepackage{dcolumn}
\usepackage{bm}
\usepackage{hyperref}
\usepackage{dsfont}
\usepackage[normalem]{ulem}
\hypersetup{
    unicode=false,          
    pdftoolbar=false,        
    pdfmenubar=true,        
    pdffitwindow=false,     
    pdfstartview={FitH},    
    pdfkeywords={Kinetic uncertainty relations, Maser heat engines, Noise-induced coherence, Current fluctuations}, 
    pdfnewwindow=true,      
    colorlinks=true,       
    linkcolor=black,          
    citecolor=blue,        
    filecolor=magenta,      
    urlcolor=blue           
}

\usepackage{xcolor}
\newcommand{\revblue}[1]{{\color{blue}#1}}
\usepackage{mathrsfs}
\usepackage{amsfonts}
\usepackage{amsmath}
\usepackage{amssymb}
\usepackage{subfigure}
\usepackage{soul}
\usepackage{physics}

\newcommand{\stkout}[1]{\ifmmode\text{\sout{\ensuremath{#1}}}\else\sout{#1}\fi}
\newcommand{\imai}{{\rm i}}

\usepackage{color}

\newcommand{\be}{\begin{equation}}
\newcommand{\bw}{\begin{widetext}}
\newcommand{\ew}{\end{widetext}}
\newcommand{\bea}{\begin{eqnarray}}
\newcommand{\bc}{\begin{center}}            
\newcommand{\ee}{\end{equation}}
\newcommand{\eea}{\end{eqnarray}}
\newcommand{\ec}{\end{center}}

\begin{document}


\title{Quantum kinetic uncertainty relations: Role of coherence in current fluctuation bounds}
 
\author{Varinder Singh}
\email{varinderkias@kias.re.kr}
\affiliation{School of Physics, Korea Institute for Advanced Study, Seoul 02455, Republic of  Korea}

\author{Euijoon Kwon}
\affiliation{School of Physics, Korea Institute for Advanced Study, Seoul 02455, Republic of  Korea}
\affiliation{Department of Physics and Astronomy \& Center for Theoretical Physics, Seoul National University, Seoul 08826, Republic of Korea} 

\author{Jae Sung Lee}
\email{jslee@kias.re.kr}
\affiliation{School of Physics, Korea Institute for Advanced Study, Seoul 02455, Republic of  Korea}

\date{\today}

\begin{abstract}
We present a comparative analysis of three quantum versions of the kinetic uncertainty relation (KUR), which incorporate coherence-dependent corrections to the classical bound. Using a three-level maser heat engine as a paradigmatic quantum thermal machine, we identify the operating regimes in which each formulation provides the tightest constraint on current fluctuations. We further identify tighter, model-specific bounds for two of these formulations, provide numerical support for both, and establish one of them analytically. Introducing ground-state degeneracy, we then demonstrate that noise-induced coherence provides a tunable resource for relocating, enhancing, or suppressing violations of the standard KUR without requiring an additional coherent drive. In particular, it enables violations in regimes where the three-level engine satisfies the classical bound. Our results clarify how distinct quantum KURs account for coherence and constrain the precision of quantum thermal machines.

\end{abstract}

\maketitle
\section{Introduction}
Quantum heat engines play a central role in the development of quantum thermodynamics~\cite{Kosloff2013,Sai2016,Sourav2021,Mahler,AlickiKosloff,DeffnerBook,Asli2020}. Beyond their theoretical significance, they are also regarded as potential components of future quantum technologies, making nanoscale energy conversion both fundamentally important and practically relevant.
The quantum regime offers features with no classical analogue, such as coherence and entanglement, that can enhance engine performance beyond classical limits \cite{Scully2001,ScullyAgarwal2003,OzgurQuantumFuel,Skrzypczyk2014,Latune2020,Lutz2014,Manzano2016,Klaers2017,Alonso2014,VOzgur2020,Nayak2026,BrijMohan2025}. At the same time, nanoscale thermal machines are inherently subject to thermal and quantum fluctuations, which can limit their stability and efficiency. Understanding the fundamental limits on these fluctuations is therefore essential for assessing the precision, stability, and reliability of quantum thermal machines \cite{PietzonkaSeifert2016,Shiraishi2016,Moreira2025,Mark2021,SinghMilburn2026,Manikandan2023,Fedorov2023,Davinder2021}.

A central framework for characterizing such limits is provided by uncertainty relations in nonequilibrium thermodynamics. In particular, the thermodynamic uncertainty relation (TUR)~\cite{SeifertTUR,Gingrich2016,Horowitz2020} establishes a trade-off between current precision and thermodynamic cost. For classical steady-state systems, it implies that suppressing current fluctuations requires a corresponding increase in entropy production, thereby placing a fundamental lower bound on the achievable fluctuations. The TUR was first derived for continuous-time Markov jump process~\cite{Gingrich2016}, and then extended to overdamped~\cite{Dechant_2018,HasegawaVu2019,JongMin2021} and underdamped~\cite{Hasegawa2019_underdamped,JSLee2021_universalTUR,Kwon2024_unified} Langevin systems, and finally to quantum systems~\cite{Hasegawa2020,VuSaito2022,Brandner2025}.

Alongside the TUR, there is a complementary family of bounds known as kinetic uncertainty relations (KURs)~\cite{Terlizzi,Garrahan2017,Maes2020,HiuraSasa}.   Rather than linking fluctuations to thermodynamic cost, KURs connect them to a system's dynamical activity, how many microscopic transitions  the system undergoes per unit time on average, denoted $A$.  Originally established for Markov jump processes with time-independent rates, it reads~\cite{Terlizzi} 
\begin{equation}
\mathcal{Q} \equiv \frac{ I^2}{A\Delta I} \le 1, \label{eq:KUR}
\end{equation}
where $I$ denotes the steady-state mean current, and $\Delta I = \lim_{t\rightarrow\infty}\langle[I(t) - I]^2\rangle\, t$ represents its scaled variance.

 \begin{figure*}    
 \begin{center}
 \includegraphics[width=0.9\textwidth]{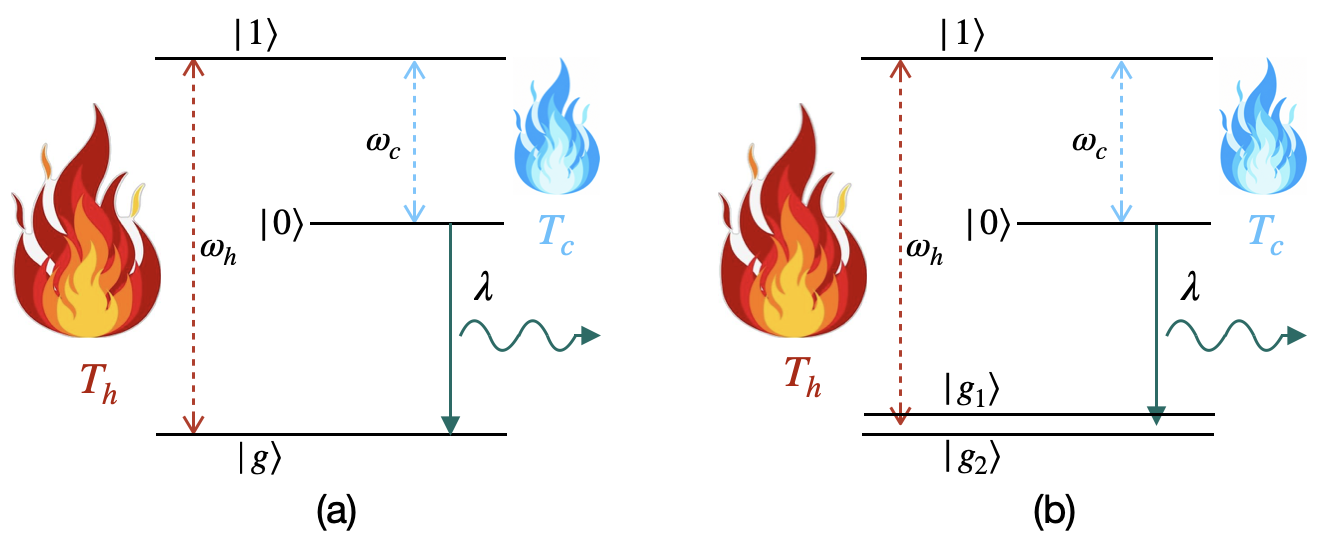}
 \end{center}
\caption{(a) Schematic of a three-level maser engine. The transition $|g\rangle \leftrightarrow |1\rangle$ (frequency $\omega_h$) is coupled to a hot reservoir at temperature $T_h$, and the transition $|0\rangle \leftrightarrow |1\rangle$ (frequency $\omega_c$) to a cold reservoir at temperature $T_c$. The transition $|0\rangle \leftrightarrow |g\rangle$ is coupled to a classical single-mode field with matter--field coupling strength $\lambda$. (b)  Schematic of a four-level maser engine with a twofold-degenerate ground state $\{|g_1\rangle, |g_2\rangle\}$; the two ground levels are drawn slightly offset for visibility.}
\label{fig:schematic}
\label{fig:models}
\end{figure*} 

In the quantum domain, however, the standard KUR in Eq.~\eqref{eq:KUR} can be violated because the classical activity bound does not fully capture the effects of coherent dynamics~\cite{Hasegawa2020,VuSaito2022,Prech2025,KwonLee2025,Vu2025PRX}. This has motivated quantum generalizations that incorporate coherence-dependent contributions into the uncertainty relation. 
Three such formulations are central to the present work. The first, introduced by Hasegawa~\cite{Hasegawa2020} and hereafter referred to as the Hasegawa KUR, supplements the dynamical activity with a quantum contribution. The $\psi$-KUR of Prech, Potts, and Landi~\cite{Prech2025} introduces a correction associated with the interplay between coherent evolution and the measured current. More recently, Kwon and Lee~\cite{KwonLee2025} derived a quantum KUR using stochastic representations of quantum-jump dynamics, which we refer to as the $\delta$-KUR. Its correction captures temporal correlations between jump noise and the fluctuating drift of the current. These formulations encode the influence of coherence in distinct ways, making their relative tightness and physical interpretation dependent on the operating regime. Comparing their performance and clarifying how they account for coherence in maser heat engines \cite{Scovil1959} are central aims of this work.

A complementary question concerns how the relevant coherence is generated. In addition to the drive-induced coherence produced by the work field, multilevel engines can support \emph{noise-induced coherence} (NIC), which arises from the interference between dissipative transitions that share a common level~\cite{scully2011,dorfman2018,Tscherbul2014,DorfmanPark}. NIC has been shown to enhance the power output of maser engines and can act as an effectively cost-free resource, since it requires no additional coherent driving \cite{scully2011,dorfman2018}. Its influence on relative power fluctuations has previously been studied in the context of the TUR~\cite{VSTUR2023,Holubec2019}. Here, we investigate its effect on violations of the standard-KUR and on the relative tightness of quantum KURs. Because NIC modifies both the steady-state coherences and the dynamical activity, it provides a natural and tunable means of probing quantum KURs and assessing whether coherence can suppress relative current fluctuations.

Guided by these two questions, in this work, we study the KUR in  three-level maser heat engine~\cite{Scovil1959,Geva1994,Geva1996,BoukobzaTannor2007,Varinder2020,Patrick2021,FeiGu2025} and its degenerate four-level extension. Building on the analysis of standard-KUR violations in a three-level maser engines~\cite{SinghKwonLeeKUR2026}, we compare the Hasegawa, $\psi$-, and $\delta$-KURs across the relevant driving regimes. We also establish a sharp, model-specific strengthening of the $\psi$-KUR and $\delta$-KUR for the three-level engine. Finally, we introduce ground-state degeneracy to generate NIC and show how it relocates, enhances, or suppresses standard-KUR violations.

The paper is organized as follows. Section~\ref{sec:model} reviews the three-level maser heat engine model. Section~\ref{sec:kur_quantum} introduces the three quantum KURs, compares them for the three-level engine, and establishes tighter $\psi$- and $\delta$-KURs. Section~\ref{sec:model2-nic} presents the degenerate four-level engine, and Sec.~\ref{sec:nic-kur} examines the effects of NIC on both the standard and quantum KURs. We conclude in Sec.~\ref{sec:conclusions}.

\section{The three-level maser heat engine Model}
\label{sec:model}

The three-level maser engine is a canonical example of a quantum heat engine. As shown in  Fig.~\ref{fig:models}(a), the setup  comprises a quantum system with three energy levels coupled simultaneously to two thermal reservoirs at temperatures $T_h$ and $T_c$, with $T_c < T_h$. The hot reservoir drives transitions between the ground state $|g\rangle$ and the upper excited state $|1\rangle$, while the cold reservoir couples the two \emph{upper} levels, facilitating relaxation between $|1\rangle$ and $|0\rangle$. The mechanism responsible for power extraction acts on the two
\emph{lower} levels: the system interacts with an external electromagnetic field
of frequency $\omega$ that drives the transition between $|g\rangle$ and $|0\rangle$. This coupling is captured by the semiclassical interaction Hamiltonian, written within the rotating-wave approximation as 
\begin{equation}
V(t) = \lambda \left( e^{-i\omega t}\, |0\rangle\langle g|
       + e^{i\omega t}\, |g\rangle\langle 0| \right),
\label{eq:Vt}
\end{equation}
where $\lambda$ is the field-matter coupling strength. Throughout this work, we restrict ourselves to the resonant case, in which the field frequency exactly matches the transition energy between the lasing levels $|g\rangle$ and $|0\rangle$, i.e., $\omega = \omega_0 - \omega_g$.

 The evolution of this three-level system, described in a frame rotating with the bare Hamiltonian $H_0 = \sum_{i=1,0,g} \omega_i\,|i\rangle\langle i|$, is governed by the Lindblad master equation:
\begin{equation}
\dot{\rho} = \mathcal{L}\rho \equiv -i[V_R, \rho] + \mathcal{L}_h[\rho] + \mathcal{L}_c[\rho].
\label{eq:lindblad}
\end{equation}
The rotating-frame interaction Hamiltonian takes the form $V_R = \lambda \left( |0\rangle\langle g| + |g\rangle\langle 0| \right)$. The dissipators $\mathcal{L}_h$ and $\mathcal{L}_c$ describe the coupling of the system to the hot and cold thermal reservoirs, respectively, and are given by
\begin{align}
\mathcal{L}_h[\rho] &= \gamma_h (n_h + 1)
  \left( \sigma_{g1}\rho\sigma_{g1}^\dagger
  - \tfrac{1}{2}\{\sigma_{g1}^\dagger\sigma_{g1}, \rho\} \right) \nonumber\\
  &\quad + \gamma_h n_h
  \left( \sigma_{g1}^\dagger\rho\sigma_{g1}
  - \tfrac{1}{2}\{\sigma_{g1}\sigma_{g1}^\dagger, \rho\} \right),
\label{eq:Lh}
\end{align}
\begin{align}
\mathcal{L}_c[\rho] &= \gamma_c (n_c + 1)
  \left( \sigma_{01}\rho\sigma_{01}^\dagger
  - \tfrac{1}{2}\{\sigma_{01}^\dagger\sigma_{01}, \rho\} \right) \nonumber\\
  &\quad + \gamma_c n_c
  \left( \sigma_{01}^\dagger\rho\sigma_{01}
  - \tfrac{1}{2}\{\sigma_{01}\sigma_{01}^\dagger, \rho\} \right),
\label{eq:Lc}
\end{align}
where $\sigma_{ij} = |i\rangle\langle j|$, so that $\sigma_{g1} = |g\rangle\langle 1|$ and $\sigma_{01} = |0\rangle\langle 1|$ are the jump operators associated with the hot and cold reservoirs, respectively. $n_h$ and $n_c$ denote the corresponding thermal occupation numbers, while $\gamma_h$ and $\gamma_c$ are the system--reservoir coupling strengths for the hot and cold baths.

The steady-state current $I$ is the average rate of forward thermodynamic cycles, $|g\rangle\to|1\rangle\to|0\rangle\to|g\rangle$, minus that of reverse cycles. A forward cycle consists of photon absorption from the hot reservoir, photon emission into the cold reservoir, and work delivery through stimulated emission.
Following the counting convention in Appendix~\ref{AppFCS}, we define $N_c(t)=N_c^{\mathrm{em}}(t)-N_c^{\mathrm{abs}}(t)$ as the net number of photons emitted into the cold reservoir during the interval $[0,t]$. The mean current is then $I=\lim_{t\to\infty}\langle N_c(t)\rangle/t$, where $\langle\cdot\rangle$ denotes an ensemble average.
In the steady state, population conservation ensures that the mean net currents along all three links of the cycle are equal. Thus, the net photon current into the cold reservoir is identical to the mean net cycle rate.

\section{Quantum versions of KUR}
\label{sec:kur_quantum}

As mentioned in the introduction, quantum systems can violate the standard KUR in Eq.~\eqref{eq:KUR}. 
We now examine the Hasegawa KUR~\cite{Hasegawa2020,VuSaito2022}, the $\psi$-KUR~\cite{Prech2025}, and the $\delta$-KUR~\cite{KwonLee2025}. The Hasegawa formulation incorporates a quantum contribution into the activity cost, whereas the $\psi$- and $\delta$-KURs retain the dynamical activity and introduce distinct correction factors. Below, we briefly review these formulations and compare how they constrain current fluctuations for the three-level maser engine.

\begin{figure}
    \centering
    \includegraphics[width=.475\textwidth]{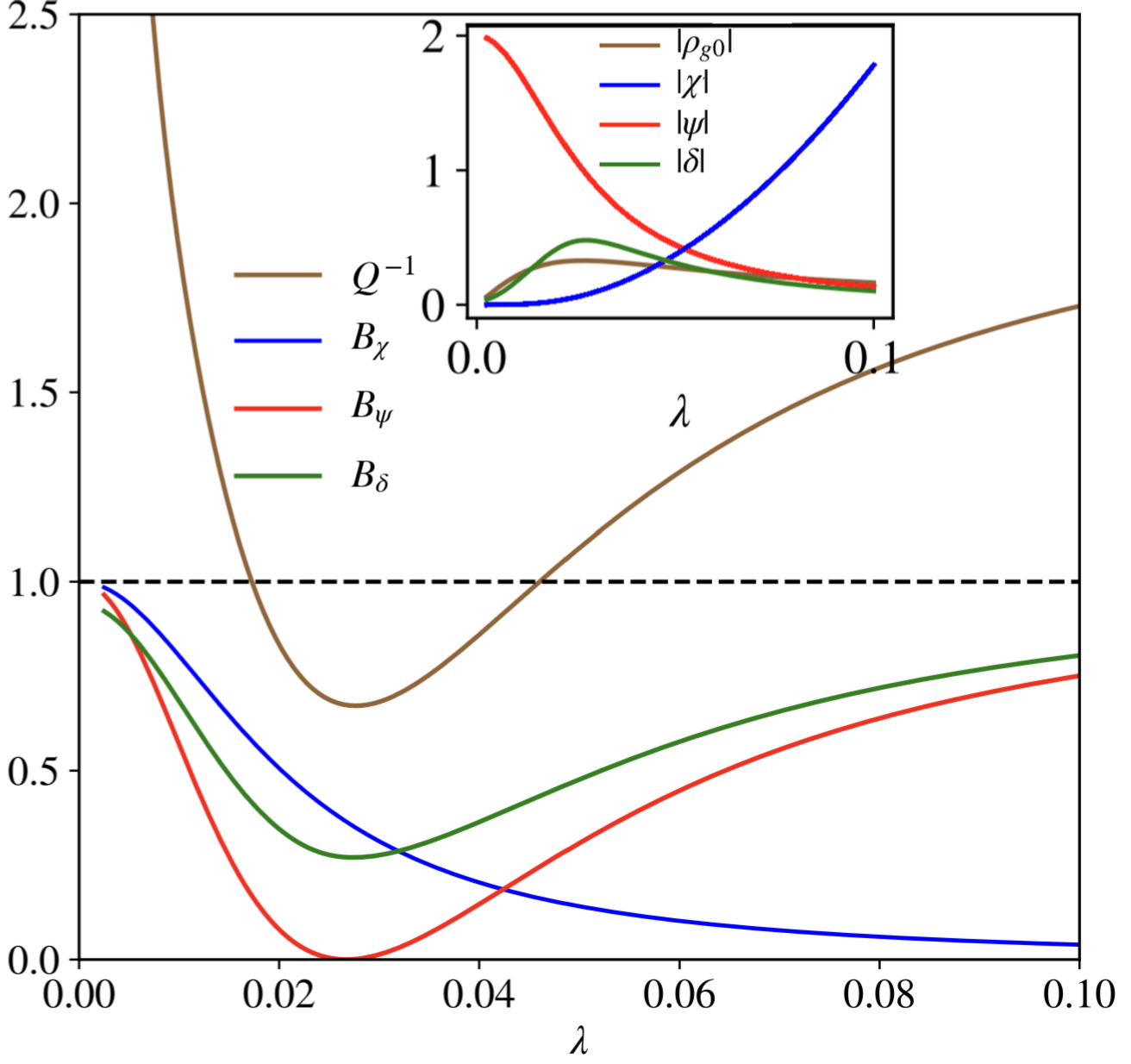}
    \caption{Quantum KURs for the three-level maser heat engine as a function of the driving strength $\lambda$. The brown curve shows $\mathcal{Q}^{-1}=A\Delta I/I^2$. The blue, red, and green curves show the lower bounds $B_\chi$ [Eq.~\eqref{HasegawaKUR}], $B_\psi$ [Eq.~\eqref{PrechKUR}], and $B_\delta$ [Eq.~\eqref{DeltaKUR}], respectively. The horizontal dashed line marks the standard-KUR threshold at unity. The inset shows the steady-state coherence $|\rho_{g0}|$ [Eq.~\eqref{cohII}] in brown, together with $|\chi|$ [Eq.~\eqref{chiclosed}], $|\psi|$ [Eq.~(\ref{eq:psi_evaluation})], and $|\delta|$ [Eq.~(\ref{eq:delta_computational})], using the same colors as their corresponding bounds in the main panel. Here, $\gamma_h=0.016$, $\gamma_c=1$, $n_h=5$, and $n_c=0.001$.
} \label{QuantumKURs}
\end{figure}

\subsection{Quantum KURs}

We begin with the first quantum extension of the KUR proposed by Hasegawa~\cite{Hasegawa2020}. This bound modifies the dynamical-activity constraint to account for quantum coherence and therefore remains applicable in regimes where the standard KUR can be violated. Accordingly, the Hasegawa KUR is written as
\begin{equation}
    \frac{\Delta I}{I^2}
    \geq \frac{1}{A+{\chi}}
    ~~~\Rightarrow~~~
    \mathcal{Q}^{-1} \equiv \frac{A\Delta I}{I^2}
    \geq \frac{A}{A+{\chi}} \equiv B_\chi,
    \label{HasegawaKUR}
\end{equation}
where ${\chi}$ quantifies the quantum correction to the dynamical-activity cost, and its explicit form is given by
\begin{equation}
\begin{aligned}
{\chi}={}&
-4\operatorname{Tr}\!\left[
{\mathcal L}_1 {\mathcal L}^{D}
{\mathcal L}_2 [\rho_{\rm ss}]\right]-4\operatorname{Tr}\!\left[
{\mathcal L}_2 {\mathcal L}^{D}
{\mathcal L}_1 [\rho_{\rm ss}]\right].
\end{aligned}
\label{QuantumCoherence}
\end{equation}
Here, ${\mathcal L}^{D}$ denotes the Drazin inverse of the Liouvillian ${\mathcal L}={\mathcal L}_1+ {\mathcal L}_2$, and ${\mathcal L}_1$ and ${\mathcal L}_2$ are the auxiliary superoperators defined in Eqs.~\eqref{EqL1L1}--\eqref{EqL1L2} of Appendix~\ref{AppFCS}.

The steady-state current and its fluctuations are evaluated using full counting statistics (FCS), as described in Appendix~\ref{AppFCS}. The resulting quantity $\mathcal{Q}$ can then be written in the closed form  \cite{SinghKwonLeeKUR2026}
\begin{equation}
\begin{aligned}
\mathcal{Q}
={}&
\frac{2(n_h-n_c)^2\gamma_c\gamma_h\lambda^2}
     {B\,B'\,H'}\\
&\times
\left[
2n_hn_c+n_h+n_c
-
\frac{
8\lambda^2(n_h-n_c)^2\gamma_c\gamma_h G'
}{
\left(4\lambda^2D' + B'C'\gamma_c\gamma_h\right)^2
}
\right]^{-1}.
\end{aligned}
\label{KURPatrick}
\end{equation}
where $B=\gamma_c(1+n_c)+\gamma_h(1+n_h)$, $B'=\gamma_c nc+\gamma_h n_h$, $C'=3n_h n_c + n_h + n_c$, $D'=\gamma_c(2+3n_c)+\gamma_h(2+3n_h)$, $G' = 8 \lambda ^2+\gamma _c \gamma _h \left(  10 n_h n_c +3n_c +3 n_h\right)+\gamma _c^2 \left(2 n_c+1\right)n_c+\gamma _h^2   \left(2 n_h+1\right)n_h$, and $H'=4\lambda^2 +  n_c n_h\gamma_c\gamma_h$.
The coherence-related contribution $\chi$ follows from Eq.~\eqref{QuantumCoherence}, evaluated using the procedure detailed in Appendix~\ref{sec:subsub_chi_psi}, and reads 
\begin{equation}
{\chi}
=
\frac{32B\lambda^2H'}
{B'\left(4\lambda^2D'+B'C'\gamma_c\gamma_h\right)}.
\label{chiclosed}
\end{equation}
Combining this result with the expression for the dynamical activity $A$ given in Eq.~\eqref{DAII}, we can evaluate the Hasegawa-KUR bound $B_\chi$.

In addition to the Hasegawa KUR, the $\psi$-KUR was proposed in recent work~\cite{Prech2025}. It is given by
\begin{equation}
\mathcal{Q}^{-1} = \frac{A\Delta I}{I^2}\geq(1+\psi)^2 \equiv B_\psi.
\label{PrechKUR}
\end{equation}
Here, $\psi$ describes a coherence-dependent correction associated with the Hamiltonian part of the dynamics. It is given by
\begin{equation}
\psi=
\frac{
\operatorname{Tr}\!\left[
\mathcal J\mathcal L^D
\mathcal H\rho_{\rm ss}\right]
}{I}.
\label{psiterm}
\end{equation}
The Hamiltonian superoperator $\mathcal H$ acts as $\mathcal H\rho=-i[H,\rho]$, while $\mathcal J$ denotes the current superoperator, with $I=\operatorname{Tr}[\mathcal J\rho_{\rm ss}]$. Thus, $\psi$ vanishes when $[H,\rho_{\rm ss}]=0$, provided $I\neq0$. 
Its value also depends on the relaxation dynamics and the measured current through $\mathcal L^D$ and $\mathcal J$, respectively, and is therefore not determined solely by the magnitude of a steady-state coherence. For the resonant maser engine, the right-hand side of Eq.~\eqref{PrechKUR} takes the closed form
\begin{equation}
B_\psi=
\frac{
\left(B'C'\gamma_c\gamma_h-4\lambda^2D'\right)^2
}{
\left(B'C'\gamma_c\gamma_h+4\lambda^2D'\right)^2
}.
\label{psiclosed}
\end{equation}
The polynomial $B'C'\gamma_c\gamma_h+4\lambda^2D'$ also appears in the expressions for ${\chi}$ and the current [see Eq.~\eqref{CurrentII}], although with different powers and prefactors. In particular, the $\psi$-KUR bound $B_\psi$ vanishes when $4\lambda^2D'=B'C'\gamma_c\gamma_h$, which corresponds to $\psi=-1$.

A third quantum extension is the $\delta$-KUR, derived by Kwon and Lee using stochastic representations of quantum-jump dynamics~\cite{KwonLee2025}. In the steady-state, long-time limit, it takes the form
\begin{equation}
\mathcal{Q}^{-1}=\frac{A\Delta I}{I^2}\geq(1+\delta)^2 \equiv B_\delta.
\label{DeltaKUR}
\end{equation}
Although this bound resembles Eq.~\eqref{PrechKUR}, the correction $\delta$ has a different construction: it captures correlations between the fluctuating drift of the measured observable and the jump noise.

To define this correction, let $\rho_c(t)$ denote the density matrix conditioned on a quantum-jump record, and let $dN_k(t)$ denote the increment that counts jumps through channel $k$. The corresponding conditional jump rate and centered increment are
\begin{equation}
\begin{aligned}
r_k(t)&=\operatorname{Tr}
\!\left[L_k\rho_c(t)L_k^\dagger\right],\\
d\widetilde N_k(t)&=dN_k(t)-r_k(t)\,dt,
\end{aligned}
\label{eq:delta_centered_jumps}
\end{equation}
respectively. 
For fixed counting weights $w_k$, the integrated current over a measurement time $\tau$ can be decomposed into fluctuating and drift
contributions as
\begin{equation}
\begin{aligned}
J_\tau
&=\sum_k w_k\int_0^\tau dN_k(t)
=J_\tau^{\rm fl}+J_\tau^{\rm dr},\\
J_\tau^{\rm fl}
&=\sum_k w_k\int_0^\tau d\widetilde N_k(t),\\
J_\tau^{\rm dr}
&=\int_0^\tau dt\sum_k w_k r_k(t).
\end{aligned}
\label{eq:delta_current_decomposition}
\end{equation}
Introducing the auxiliary observable
\begin{equation}
Z_\tau=\sum_k\int_0^\tau d\widetilde N_k(t),
\label{eq:delta_auxiliary}
\end{equation}
the correction entering Eq.~\eqref{DeltaKUR} is defined as
\begin{equation}
\delta = 
\lim_{\tau\to\infty}
\frac{\left\langle J_\tau^{\rm dr}Z_\tau\right\rangle}
     {\left\langle J_\tau\right\rangle}.
\label{deltaterm}
\end{equation}
Here, angular brackets denote averages over quantum-jump trajectories. For stationary dynamics, $\langle Z_\tau\rangle=0$, $\langle Z_\tau^2\rangle=A\tau$, and $\langle J_\tau^{\rm fl}Z_\tau\rangle =\langle J_\tau\rangle$.
Applying the Cauchy--Schwarz inequality to $\langle J_\tau Z_\tau \rangle$ and subsequently taking the long-time limit yields Eq.~\eqref{DeltaKUR}.

The correction $\delta$ therefore contains information about temporal correlations in the conditioned dynamics and can remain nonzero in a quantum steady state. For the three-level maser, the thermal jump operators reset the system to definite levels, placing the dynamics within the quantum reset class discussed in Ref.~\cite{KwonLee2025}. Thus, $\delta$ characterizes quantum dynamical effects in a manner distinct from both the additional activity cost $\chi$ and the Hamiltonian correction $\psi$. An explicit procedure for evaluating $\delta$ is given in Appendix~\ref{app:delta_evaluation}.

\begin{figure*}
    \centering
    \includegraphics[width=\linewidth]{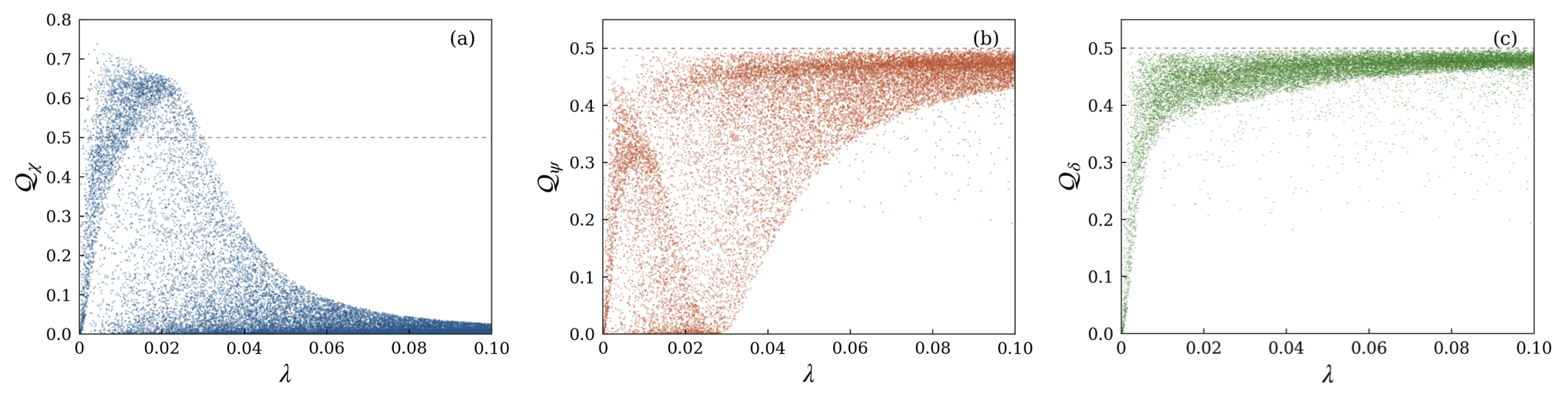}
\caption{   
Scatter plots of (a) $\mathcal{Q}_{\chi}$, (b) $\mathcal{Q}_{\psi}$, and (c) $\mathcal{Q}_{\delta}$, defined in Eq.~\eqref{bothKUR}, as functions of the driving strength $\lambda$.
Each point corresponds to one of $2\times10^{4}$ independently sampled parameter sets, drawn uniformly from $n_c\in[0,0.001]$, $n_h\in[4.99,5]$, $\gamma_c\in[0.9,1]$, $\gamma_h\in[10^{-4},0.017]$, and $\lambda\in[0,0.1]$. The same parameter sets are used in all three panels, and the horizontal dashed lines indicate $1/2$. 
While some sampled values of $\mathcal{Q}_{\chi}$ exceed $1/2$, all sampled values of $\mathcal{Q}_{\psi}$ and $\mathcal{Q}_{\delta}$ remain below this threshold. This is consistent with the analytically established $\psi$ bound and provides numerical support for the conjectured $\delta$ bound in Eq.~\eqref{eq:tighter_psi_kur}.}
    \label{hist}
\end{figure*}

To compare the quantum KURs, we plot Eqs.~\eqref{HasegawaKUR}, \eqref{PrechKUR}, and \eqref{DeltaKUR} in Fig.~\ref{QuantumKURs} for the three-level maser heat engine. The common left-hand side, $\mathcal{Q}^{-1} = A\Delta I/I^2$, is shown by the brown curve.
The blue, red, and green curves represent $B_\chi$, $B_\psi$, and $B_\delta$, respectively. Since all three inequalities share the same left-hand side, a larger value of the right-hand side corresponds to a tighter lower bound on $\mathcal{Q}^{-1}$. The brown curve falls below unity over an intermediate range of $\lambda$, explicitly demonstrating a violation of the standard KUR, while remaining above all three quantum bounds.
 
At weak driving, $B_\chi$ is the tightest of the three bounds. As $\lambda$ increases, $B_\chi$ decreases, whereas $B_\psi$ and $B_\delta$ exhibit nonmonotonic behavior. The bound $B_\psi$ reaches zero near $\lambda\simeq0.027$, where $\psi=-1$, before increasing again. By contrast, $B_\delta$ retains a finite minimum and overtakes $B_\chi$ near $\lambda\simeq0.033$. Beyond this crossover, $B_\delta$ provides the tightest of the three displayed bounds. The bound $B_\psi$ also eventually exceeds the Hasegawa bound, but remains below $B_\delta$ in this stronger-driving regime. The same qualitative trend is observed across the parameter regimes explored: $B_\chi$ is tighter at weak driving, whereas $B_\delta$ becomes the tightest of the three at stronger driving strengths. The crossover points depend on the system parameters.

The inset compares the steady-state coherence $|\rho_{g0}|$ [Eq.~\eqref{cohII}] with the correction terms $|\chi|$ [Eq.~\eqref{chiclosed}], $|\psi|$ [Eq.~(\ref{eq:psi_evaluation})], and $|\delta|$ [Eq.~(\ref{eq:delta_computational})], using the same colors for the correction terms as for their corresponding bounds in the main panel; the coherence is shown in brown. As $\lambda$ increases, both $|\rho_{g0}|$ and $|\delta|$ initially increase, reach a maximum, and then decrease, whereas $|\psi|$ decreases and $|\chi|$ increases monotonically over the displayed range. Thus, $|\delta|$ reflects the nonmonotonic behavior of the steady-state coherence, while $|\psi|$ follows its decreasing trend only at stronger driving.

These differences help interpret the relative tightness of the bounds. At larger driving strengths, $\chi$ continues to increase despite the decrease in the steady-state coherence, leading to a progressively looser $B_\chi$ bound. In contrast, the decreasing magnitudes of the negative corrections $\psi$ and $\delta$ allow their corresponding bounds to increase toward the classical threshold. Their closer correspondence with the coherence in this regime is accompanied by tighter constraint on current fluctuations, with the $\delta$-KUR providing the tightest bound among the three formulations.  Although these corrections are not themselves direct measures of coherence, their behavior suggests that the $\psi$- and $\delta$-KURs more effectively capture the contribution of coherence to the uncertainty relation in this regime.

\subsection{Tighter quantum KURs for the three-level maser engine}

To investigate whether the three quantum KURs admit tighter bounds for the three-level maser engine, we examine the corresponding dimensionless quantifiers as functions of the driving strength $\lambda$, as shown in Fig.~\ref{hist}. We randomly sample the system parameters over the specified intervals and evaluate
\begin{equation}
\begin{aligned}
\mathcal{Q}_{\chi}
&\equiv \mathcal{Q}B_\chi
= \frac{I^2}{\Delta I(A+\chi)},\\
\mathcal{Q}_{\psi}
&\equiv \mathcal{Q}B_\psi
= \frac{I^2}{A\Delta I}(1+\psi)^2,\\
\mathcal{Q}_{\delta}
&\equiv \mathcal{Q}B_\delta
= \frac{I^2}{A\Delta I}(1+\delta)^2.
\end{aligned}
\label{bothKUR}
\end{equation}
Each point represents one sampled parameter set. The original quantum KURs in Eqs.~\eqref{HasegawaKUR}, \eqref{PrechKUR}, and \eqref{DeltaKUR}, require all three quantities to satisfy $\mathcal{Q}_{\alpha}\leq 1$, where $\alpha\in\{\chi,\psi,\delta\}$. 

The scatter plots reveal a clear distinction among the three formulations. The sampled values of $\mathcal{Q}_{\chi}$ exceed $1/2$ at weak to intermediate driving and become small at stronger driving. In contrast, both $\mathcal{Q}_{\psi}$ and $\mathcal{Q}_{\delta}$ remain below $1/2$ throughout the sampled parameter region, with many values approaching this threshold at larger $\lambda$. Moreover, $\mathcal{Q}_{\delta}$ is concentrated near $1/2$ over a broader range of driving strengths, whereas $\mathcal{Q}_{\psi}$ exhibits a wider spread at weak to intermediate driving. These numerical observations suggest the tighter bounds
\begin{equation}
\mathcal{Q}_{\psi}\leq\frac{1}{2},
\qquad
\mathcal{Q}_{\delta}\leq\frac{1}{2}.
\label{eq:tighter_psi_kur}
\end{equation}

For the $\psi$-KUR, we prove this bound analytically  in Appendix~\ref{app:tighter_psi_kur}. For the $\delta$-KUR, the bound is supported by numerical sampling, while its validity over the full parameter domain remains conjectural. Each inequality strengthens the corresponding original KUR by a factor of two, equivalently requiring $\mathcal{Q}^{-1}\geq 2B_\alpha$ for $\alpha\in\{\psi,\delta\}$, within the respective scope established above.

Finally, the constant $1/2$ in the strengthened $\psi$-KUR of Eq.~\eqref{eq:tighter_psi_kur} is sharp. Although the inequality is strict for every finite set of strictly positive parameters, the value $1/2$ can be approached arbitrarily closely. Consider the parameter family
\begin{equation}
\begin{gathered}
n_c=\varepsilon^2,\qquad n_h=1,\qquad
\gamma_c=\lambda=\Gamma,\qquad
\gamma_h=\varepsilon\Gamma,
\end{gathered}
\label{eq:psi_sharp_family}
\end{equation}
where $\Gamma>0$ is fixed and $0<\varepsilon<1$. Along this family, $\mathcal{Q}_{\psi}\to1/2$ as $\varepsilon\to0^+$. Thus, no smaller parameter-independent constant can replace $1/2$ in the strengthened $\psi$-KUR over the stated parameter domain.

\section{Four-level degenerate version of maser heat engine}
\label{sec:model2-nic}
We now consider the degenerate four-level counterpart of the three-level system, as depicted in Fig.~\ref{fig:models}(b). Here the degeneracy lies in the \emph{ground} manifold: the single ground state $|g\rangle$ of the three-level model is replaced by a pair of degenerate states, $|g_1\rangle$ and $|g_2\rangle$. 
The hot reservoir drives both transitions $|g_1\rangle\leftrightarrow|1\rangle$ and $|g_2\rangle\leftrightarrow|1\rangle$, with Wigner--Weisskopf rates $\gamma_{h1}$ and $\gamma_{h2}$. 
Consequently, the common upper level $|1\rangle$ is coupled to two indistinguishable decay channels, $|1\rangle\to|g_1\rangle$ and $|1\rangle\to|g_2\rangle$. Interference between these channels gives rise to noise-induced coherence.

 For the four-level system, the bare Hamiltonian is $
H_0 = \sum_{i=1,0,g_1,g_2}\omega_i\,|i\rangle\langle i|$, where $\omega_{g_1}=\omega_{g_2}\equiv\omega_g$. Within the rotating-wave approximation, the semiclassical system--field interaction couples the degenerate ground-state manifold to $|0\rangle$, $ V(t) = \lambda\big[e^{-i\omega t}\big(|0\rangle\langle g_1|+|0\rangle\langle g_2|\big) + e^{i\omega t}\big(|g_1\rangle\langle 0|+|g_2\rangle\langle 0|\big)\big].$ 
We work at resonance with the lasing transition, $\omega = \omega_0-\omega_g = \omega_h-\omega_c$, where $\omega_h=\omega_1-\omega_g$ and $\omega_c=\omega_1-\omega_0$ are the hot and cold transition frequencies, respectively. In the rotating frame, the dynamics is still governed by Eq.~\eqref{eq:lindblad}, with modified forms of $V_R$ and $\mathcal{L}_h$. The coherent coupling becomes $V_R = \lambda\big(|0\rangle\langle g_1|+|0\rangle\langle g_2|+\mathrm{H.c.}\big)$ while the hot-bath dissipator takes the form
\begin{align}
\mathcal L_h[\rho] =
\sum_{i,j=1}^{2} \gamma_{ij}^{(h)}
&\left[ (n_h+1) \left( A_i\rho A_j^\dagger -\frac{1}{2}\{A_j^\dagger A_i,\rho\} \right) \right.  \nonumber\\
& \left. + n_h \left( A_i^\dagger\rho A_j -\frac{1}{2}\{A_j A_i^\dagger,\rho\} \right) \right],
\end{align}
where the hot-transition operators are
$ A_i \equiv |g_i\rangle\langle 1|$ ($i=1,2$),
and the hot-reservoir rate matrix is
\begin{equation}
\gamma_{ij}^{(h)} =
\begin{pmatrix}
\gamma_{h1} & \Gamma\cos\theta\\
\Gamma\cos\theta & \gamma_{h2}
\end{pmatrix}_{ij},
\qquad
\Gamma=\sqrt{\gamma_{h1}\gamma_{h2}}.
\end{equation}
Here, $\theta$ denotes the angle between the transition dipole moments associated with  $|1\rangle\to|g_1\rangle$ and $|1\rangle\to|g_2\rangle$. We set $\gamma_{h1}=\gamma_{h2}=\gamma_h$, and define the noise-induced-coherence parameter $p\equiv\cos\theta\in(-1,1)$.

\section{Effect of noise-induced coherence on  KUR}
\label{sec:nic-kur}
Here, we compare the KURs for the three-level and four-level maser heat engines and examine how noise-induced coherence enhances or suppresses current fluctuations.

\subsection{Comparison with the Standard KUR}
\label{sec:va}

\begin{figure}
    \centering
    \includegraphics[width=1\linewidth]{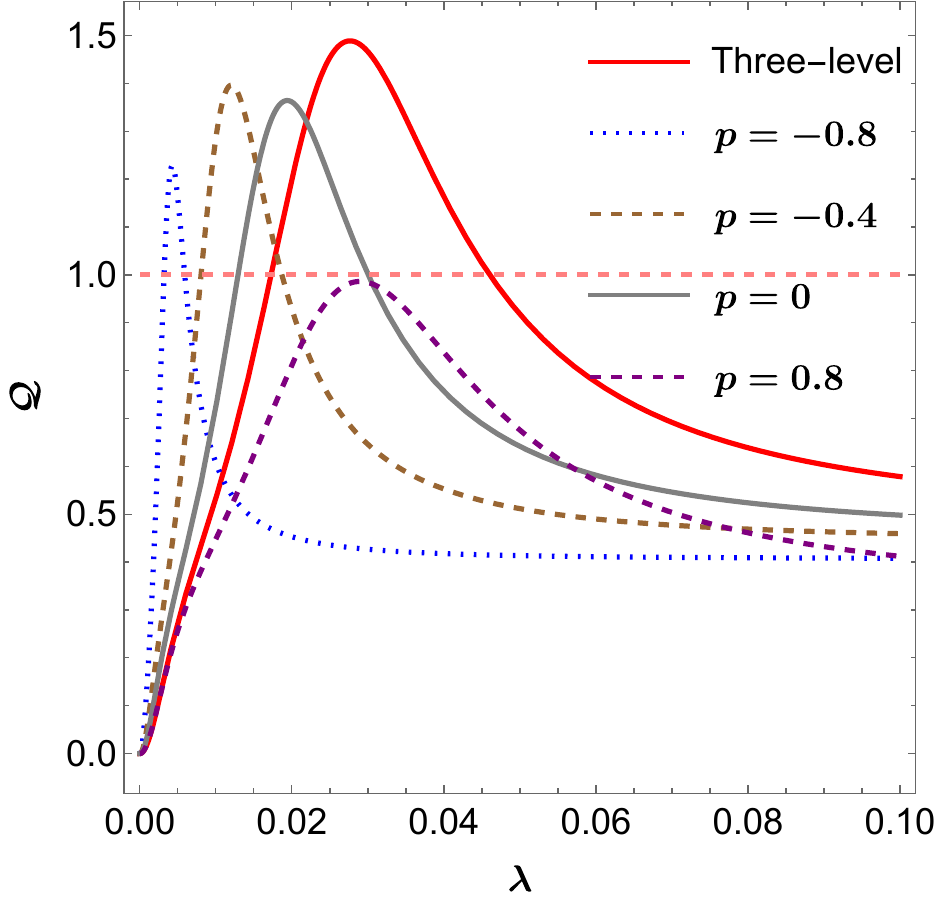}
    \caption{ 
Plots of $\mathcal{Q}$  versus the matter--field coupling $\lambda$ for the three-level and  degenerate four-level engines. Values $\mathcal{Q}>1$ indicate a violation of the standard KUR. The solid red curve represents the nondegenerate three-level engine, while the dotted blue, dashed brown, solid gray, and dashed purple curves correspond to the degenerate four-level engine with $p=-0.8$, $-0.4$, $0$, and $0.8$, respectively. The parameters correspond to a regime of pronounced KUR violation: the three-level engine reaches the highest peak ($\mathcal{Q}\approx1.5$), negative $p$ shifts and sharpens the violation toward smaller $\lambda$, and positive $p$ suppresses the violation. Here, $\gamma_h=0.016$, $\gamma_c=1$, $n_h=5$, and $n_c=0.001$.
}
    \label{fig:kur_max}
\end{figure}

As in Sec.~\ref{sec:kur_quantum}, we evaluate $\mathcal{Q} = I^2/(A\Delta I)$ for the degenerate four-level maser engine considered here. A value $\mathcal{Q}>1$ signals a violation of the classical KUR. The closed-form expression for $\mathcal{Q}$ [see Eq.~\eqref{eq:Qdeg_explicit}] is obtained using full counting statistics, as detailed in Appendix~\ref{AppFCS}.

In Fig.~\ref{fig:kur_max} we plot $\mathcal{Q}$  versus the matter--field coupling $\lambda$ at $n_c=0.001$, a regime in which pronounced violations of the standard KUR occur. The solid red curve corresponds to the nondegenerate three-level engine (3LS), while the remaining curves represent the degenerate four-level engine for several values of the noise-induced-coherence parameter $p$. In all cases, $\mathcal{Q}$ rises from zero, reaches a maximum at an intermediate coupling strength, and decreases at large $\lambda$. 
Among the curves shown, the three-level engine attains the largest maximum value, $\mathcal{Q}\approx 1.5$. This advantage, however, is not uniform across $\lambda$. Negative $p$ shifts the peak of the four-level engine toward smaller $\lambda$ and makes it sharper; for example, for $p=-0.8$, the peak occurs near $\lambda\approx 0.005$. Consequently, in the weak-coupling region, the four-level curves can exceed the three-level curve, yielding a larger value of $\mathcal{Q}$. By contrast, for positive $p$ (e.g., $p=0.8$), the peak is suppressed below unity, restoring the standard KUR. Thus, which engine exhibits the stronger violation of the standard KUR depends on $\lambda$, and noise-induced coherence can relocate, enhance, or suppress the violation depending on the sign and magnitude of $p$.

\begin{figure}
    \centering
    \includegraphics[width=1\linewidth]{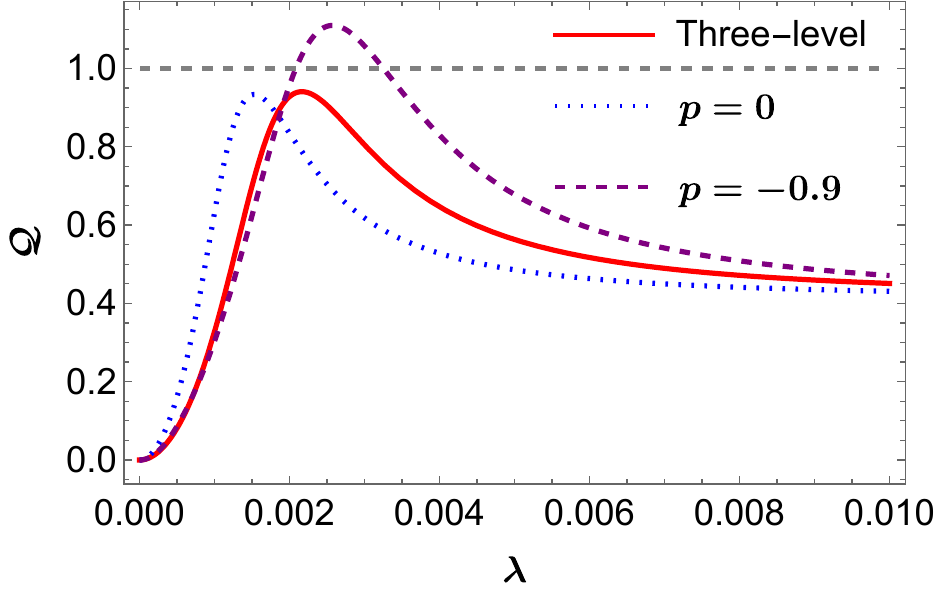}
    \caption{ $\mathcal{Q}$ as a function of $\lambda$ for both three- and four-level maser engines. The horizontal dashed line marks the threshold $\mathcal{Q}=1$. The solid red curve represents the nondegenerate three-level engine, the dotted blue curve the degenerate four-level engine with $p=0$, and the dashed purple curve the degenerate four-level engine with $p=-0.9$. The three-level and $p=0$ curves remain below the threshold, whereas a strongly negative value of the noise-induced-coherence parameter ($p=-0.9$) raises $\mathcal{Q}$ above the three-level curve near its peak and slightly above unity, demonstrating a standard-KUR violation induced by noise-induced coherence. Here, $\gamma_h=0.001$, while the remaining parameters are the same as those in Fig.~\ref{fig:kur_max}. }
    \label{fig:kur_supp}
\end{figure}

A complementary regime is shown in Fig.~\ref{fig:kur_supp}, obtained for a smaller value of $\gamma_h$ than in Fig.~\ref{fig:kur_max}, with all other parameters held fixed.  Here, the ordering of the peak values is reversed, with the four-level engine attaining a larger maximum $\mathcal{Q}$ than its three-level counterpart.
Both the three-level engine (solid red) and the $p=0$ four-level engine (dotted blue) remain below unity and therefore satisfy the standard KUR. By contrast, a strongly negative value of the noise-induced-coherence parameter ($p=-0.9$, dashed purple) raises $\mathcal{Q}$ above the three-level curve and pushes its maximum above unity, to $\mathcal{Q}\approx 1.12$. 
Thus, noise-induced coherence allows the four-level engine to violate the standard KUR in a regime where the three-level engine still satisfies it. This demonstrates that noise-induced coherence can further suppress the activity-weighted relative current fluctuations compared with the three-level engine. Noise-induced coherence thus serves as a resource for this improvement without requiring an additional coherent drive, arising instead from interference between reservoir-induced transitions involving the degenerate ground states.

\subsection{Comparison with the Quantum KURs}

\begin{figure}
    \centering
    \includegraphics[width=\linewidth]{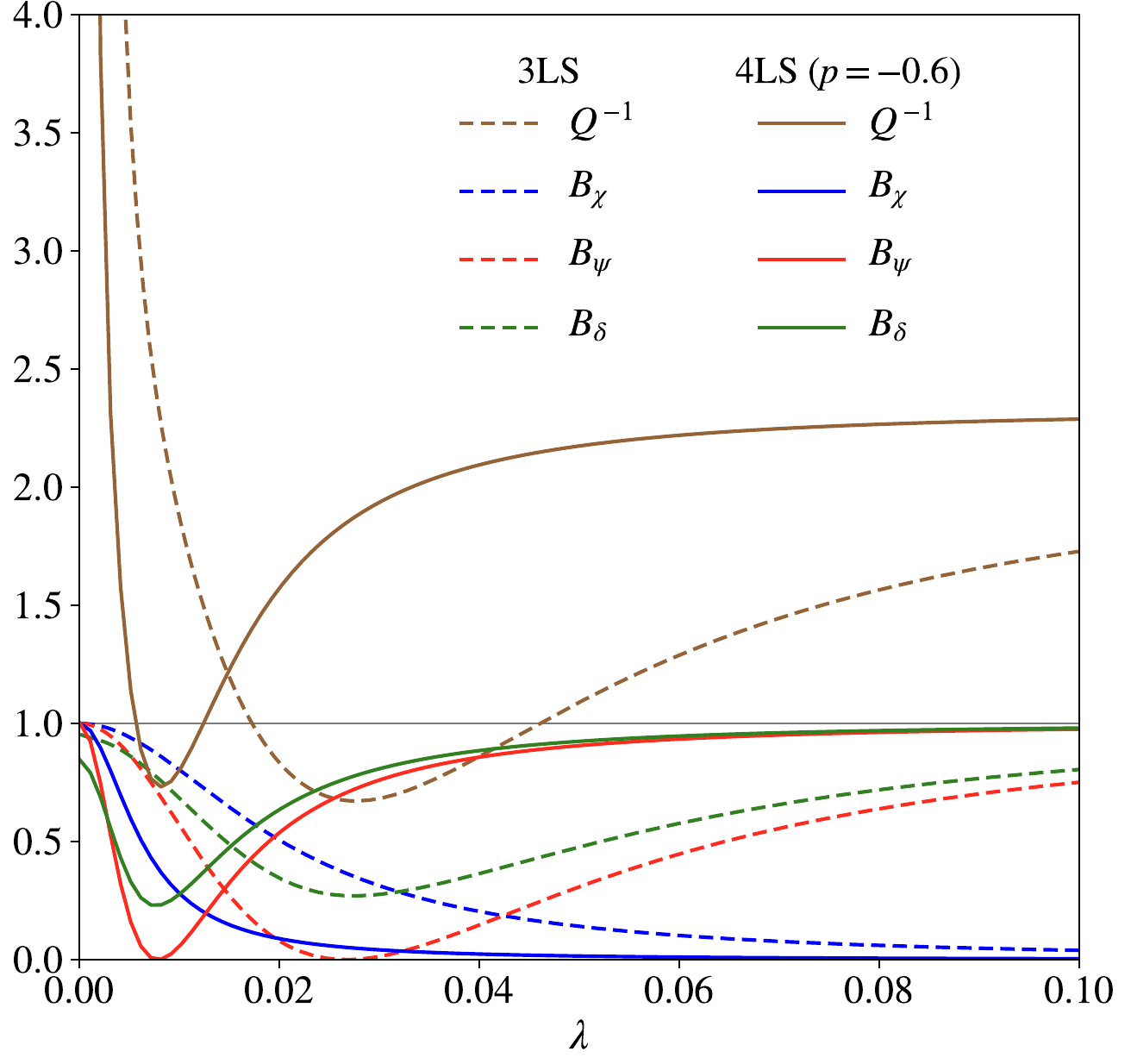}
    \caption{Quantum KURs for the three-level engine (3LS, dashed curves) and the degenerate four-level engine at $p=-0.6$ (4LS, solid curves). The brown curves show $\mathcal{Q}^{-1}$, while the blue, red, and green curves show the corresponding lower bounds $B_\chi$ [Eq.~\eqref{HasegawaKUR}], $B_\psi\equiv(1+\psi)^2$ [Eq.~\eqref{PrechKUR}], and $B_\delta\equiv(1+\delta)^2$ [Eq.~\eqref{DeltaKUR}], respectively. The horizontal gray line marks the standard-KUR threshold, $\mathcal{Q}^{-1}=1$. The three-level engine violates the standard KUR at intermediate $\lambda$, whereas the four-level engine violates it within a narrow window at smaller $\lambda$. All three quantum bounds remain below the corresponding $\mathcal{Q}^{-1}$ throughout the displayed range. For both engines, the Hasegawa bound is the tightest at weak driving, while the $\delta$-KUR becomes the tightest among the three original quantum formulations at larger driving strengths. The parameters are same as those used in Fig. 2.
}
    \label{fig:quantum-nic}
\end{figure}

In Sec.~\ref{sec:va}, we found that noise-induced coherence can drive the degenerate four-level engine into violation of the standard KUR. We now examine how the three quantum KURs introduced in Sec.~\ref{sec:kur_quantum} constrain current fluctuations  when noise-induced coherence is present. Specifically, we evaluate the common left-hand side $\mathcal{Q}^{-1} = A\Delta I/I^2$ together with the Hasegawa bound $B_\chi$ [Eq.~\eqref{HasegawaKUR}], the $\psi$-KUR bound $B_\psi$ [Eq.~\eqref{PrechKUR}], and the $\delta$-KUR bound $B_\delta$ [Eq.~\eqref{DeltaKUR}]. The results are shown in Fig.~\ref{fig:quantum-nic} as functions of $\lambda$. Dashed curves correspond to the three-level engine, whereas solid curves represent the degenerate four-level engine at $p=-0.6$. The brown curves show $\mathcal{Q}^{-1}$, while the blue, red, and green curves represent $B_\chi$, $B_\psi$, and $B_\delta$, respectively.

Consider first $\mathcal{Q}^{-1}$. For the three-level engine (dashed brown), $\mathcal{Q}^{-1}$ falls below unity over an intermediate window of $\lambda$, reaching a minimum of about $0.65$, corresponding to the standard-KUR violation already discussed. The degenerate four-level engine (solid brown) behaves differently: $\mathcal{Q}^{-1}$ drops below unity only within a narrow window at small $\lambda$, before rising steeply and saturating well above unity at large $\lambda$. Thus, for the parameters shown, the two systems violate the standard KUR over disjoint driving intervals: the three-level engine at moderate $\lambda$ and the degenerate four-level engine at small $\lambda$. This is consistent with the reshaping of the violation by noise-induced coherence found in Sec.~\ref{sec:va}.

For both engines, all three quantum KURs remain satisfied throughout the displayed range, including the region where the standard KUR is violated. Their relative tightness follows the same general trend discussed in Sec.~\ref{sec:kur_quantum}: the Hasegawa bound is tightest at weak driving, whereas the $\delta$-KUR provides the strongest constraint among the three original quantum formulations at larger $\lambda$. Noise-induced coherence shifts the crossover scales without changing this qualitative ordering. In particular, at strong driving, the four-level $\psi$- and $\delta$-KUR bounds approach unity more closely than their three-level counterparts, but remain appreciably below the four-level value of $\mathcal{Q}^{-1}$.

\section{Conclusions}
\label{sec:conclusions}

We have investigated the KUR in the three-level  maser heat engine and its degenerate four-level extension, with two complementary goals: to compare recently proposed quantum generalizations of the KUR, and to determine how noise-induced coherence influences the current fluctuations of the engine.

Our comparison shows that the $\delta$-KUR provides the tightest bound among the three original formulations over most of the driving range examined, while the Hasegawa KUR is tighter at weak driving. Beyond this comparison, we obtain the tighter bounds $\mathcal Q_{\psi,\delta}\leq1/2$ for the three-level engine, supported analytically or numerically. Both bounds can be nearly saturated, whereas $\mathcal Q_\chi$ can exceed $1/2$, so the same tighter bound does not hold for the Hasegawa KUR.

Extending the analysis to the four-level engine, we show that noise-induced coherence provides a tunable resource that can relocate, enhance, or suppress standard-KUR violations. In particular, we identify a regime where it enables the four-level engine to violate the standard KUR while its three-level counterpart still satisfies it. This improvement in current precision relative to dynamical activity requires no additional coherent drive, since noise-induced coherence arises from interference between reservoir-induced transitions involving the degenerate ground states.

Taken together, our findings show that all three quantum KURs remain valid with noise-induced coherence, which shifts their crossover scales without changing the qualitative trends in their relative tightness. More broadly, these results highlight how the tightness of quantum KURs depends on both the form of the quantum correction and the underlying engine dynamics.

 \section{Acknowledgements} 
 This research was supported by individual KIAS Grant Nos. PG064902 (J.S.L.), QP10301 (E.K.)  and   PG096801 (V.S.) atthe Korea Institute for Advanced Study.

\onecolumngrid

\appendix\section{Density matrix equations for three-level and four-level maser heat  engines}
We give the density-matrix evolution equations for both maser configurations below.

\subsection{Three-level engine}

For the three-level system in Fig.~1(a), the relevant density-matrix elements obey the equations below. The coherences $\rho_{01}$ and $\rho_{g1}$, together with their Hermitian conjugates, decay to zero in the steady state and are therefore omitted below:
 \begin{eqnarray}
\dot\rho_{11} &= &\gamma_h n_h\rho_{gg} +\gamma_c n_c\rho_{00}-[\gamma_h(n_h+1)+\gamma_c(n_c+1)] \rho_{11}, \nonumber
\\
\vspace{1mm} 
\\
\dot\rho_{00} &=& \gamma_c (n_c+1)\rho_{11} - \gamma_c n_c \rho_{00} + \imai \lambda (\rho_{0g}-\rho_{g0})\label{Eqrho00} ,
\\
\dot\rho_{gg} &=& \gamma_h (n_h+1)\rho_{11} - \gamma_h n_h\rho_{gg} -\imai\lambda (\rho_{0g}-\rho_{g0})\label{Eqrholl}\,,
\\
\dot\rho_{g0} &=&  \imai\lambda(\rho_{00}-\rho_{gg})-\frac{1}{2}(\gamma_h n_h+\gamma_c n_c) \rho_{g0}\,, 
\\
\dot\rho_{0g} &=& \dot\rho^*_{g0}  \label{Eqrhoul}
\end{eqnarray}

Solving these equations in the stationary regime yields
 \begin{eqnarray}
   \rho_{g0} &=& \frac{2i(n_c-n_h)\gamma_h\gamma_c\lambda}
    {4\lambda^2[\gamma_h(3n_h+2)+\gamma_c(3n_c+2)] + (3n_h n_c+n_h+n_c)(\gamma_h n_h+\gamma_c n_c)\gamma_h\gamma_c
} , \label{cohII}
\\
   \rho_{gg} &=& \frac{n_c(1+n_h)(\gamma_c n_c+\gamma_h n_h) \gamma_h\gamma_c + 4\lambda^2\left[\gamma_c(1+n_c)+\gamma_h(1+n_h)\right]}
    {4\lambda^2[\gamma_h(3n_h+2)+\gamma_c(3n_c+2)] + (3n_h n_c+n_h+n_c)(\gamma_h n_h+\gamma_c n_c)\gamma_h\gamma_c
} , \label{rhoggII}
\\
   \rho_{00} &=& \frac{n_h(1+n_c)(\gamma_c n_c+\gamma_h n_h) \gamma_h\gamma_c + 4\lambda^2\left[\gamma_c(1+n_c)+\gamma_h(1+n_h)\right]}
    {4\lambda^2[\gamma_h(3n_h+2)+\gamma_c(3n_c+2)] + (3n_h n_c+n_h+n_c)(\gamma_h n_h+\gamma_c n_c)\gamma_h\gamma_c
} , \label{rho00II}
\\
   \rho_{11} &=& \frac{(\gamma_c n_c+\gamma_h n_h) \left[4\lambda^2+ n_c n_h\gamma_c\gamma_h\right]}
    {4\lambda^2[\gamma_h(3n_h+2)+\gamma_c(3n_c+2)] + (3n_h n_c+n_h+n_c)(\gamma_h n_h+\gamma_c n_c)\gamma_h\gamma_c
} . \label{rho11II}
\end{eqnarray}

\subsection{Four-level degenerate heat engine}

In the four-level degenerate variant, the populations ($\rho_{11},\rho_{00},\rho_{g_1g_1},\rho_{g_2g_2}$), the noise-induced coherence $\rho_{g_1g_2}$, and the lasing coherences $\rho_{0g_1},\rho_{0g_2}$ obey the following closed set of equations. The coherences $\rho_{1g_k}$ and $\rho_{10}$ are not directly sourced and decay to zero in the steady state:
\begin{align}
\dot\rho_{11} &= -\big[2\gamma_h(n_h+1)+\gamma_c(n_c+1)\big]\rho_{11}
  + \gamma_h n_h\big(\rho_{g_1g_1}+\rho_{g_2g_2}\big)
  + \gamma_c n_c\,\rho_{00}
  + p\gamma_h n_h\big(\rho_{g_1g_2}+\rho_{g_2g_1}\big),\\
\dot\rho_{00} &= i\lambda\big(\rho_{0g_1}+\rho_{0g_2}-\rho_{g_10}-\rho_{g_20}\big)
  + \gamma_c(n_c+1)\rho_{11} - \gamma_c n_c\,\rho_{00},\\
\dot\rho_{g_1g_1} &= i\lambda\big(\rho_{g_10}-\rho_{0g_1}\big)
  + \gamma_h(n_h+1)\rho_{11} - \gamma_h n_h\,\rho_{g_1g_1}
  - \tfrac12 p\gamma_h n_h\big(\rho_{g_1g_2}+\rho_{g_2g_1}\big),\\
\dot\rho_{g_2g_2} &= i\lambda\big(\rho_{g_20}-\rho_{0g_2}\big)
  + \gamma_h(n_h+1)\rho_{11} - \gamma_h n_h\,\rho_{g_2g_2}
  - \tfrac12 p\gamma_h n_h\big(\rho_{g_1g_2}+\rho_{g_2g_1}\big),\\
\dot\rho_{g_1g_2} &= -i\lambda\big(\rho_{0g_2}-\rho_{g_10}\big)
  - \gamma_h n_h\,\rho_{g_1g_2} + p\gamma_h(n_h+1)\rho_{11}
  - \tfrac12 p\gamma_h n_h\big(\rho_{g_1g_1}+\rho_{g_2g_2}\big),\\
\dot\rho_{0g_1} &= i\lambda\big(\rho_{00}-\rho_{g_1g_1}-\rho_{g_2g_1}\big)
  - \tfrac12\big(\gamma_c n_c+\gamma_h n_h\big)\rho_{0g_1}
  - \tfrac12 p\gamma_h n_h\,\rho_{0g_2},\\
\dot\rho_{0g_2} &= i\lambda\big(\rho_{00}-\rho_{g_2g_2}-\rho_{g_1g_2}\big)
  - \tfrac12\big(\gamma_c n_c+\gamma_h n_h\big)\rho_{0g_2}
  - \tfrac12 p\gamma_h n_h\,\rho_{0g_1},
\end{align}
together with the trace condition $\rho_{11}+\rho_{00}+\rho_{g_1g_1}+\rho_{g_2g_2}=1$ and the Hermiticity relations $\rho_{g_2g_1}=\rho_{g_1g_2}^{*}$, $\rho_{g_k0}=\rho_{0g_k}^{*}$.

\section{Full Counting Statistics}
\label{AppFCS}

For the three-level maser, a forward thermodynamic cycle consists of photon absorption from the hot reservoir, photon emission into the cold reservoir, and work extraction through stimulated emission. The net current $I$ is the average rate of forward cycles, $\ket{g}\rightarrow\ket{1}\rightarrow\ket{0}\rightarrow\ket{g}$, minus that of reverse cycles. The power output is therefore the product of the net current and the energy transferred to the work field per forward cycle, yielding
\begin{equation}
P=(\omega_h-\omega_c)I.
\label{aux1}
\end{equation}
Similarly, the scaled power variance is given by \cite{VSTUR2023,Patrick2021}
\begin{equation}
\Delta P=(\omega_h-\omega_c)^2\Delta I.
\label{aux2}
\end{equation}
Using Eqs.~\eqref{aux1} and \eqref{aux2}, the KUR in Eq.~\eqref{eq:KUR} can be written as
\begin{equation}
\frac{A\Delta I}{I^2}\geq1.
\end{equation}
We compute the average current $I$ and its variance $\Delta I$ using full counting statistics (FCS), which characterizes transport in open quantum systems by introducing counting fields into the master equation.
We first detail the calculation of the current statistics and the three quantum corrections for the three-level engine. We then present the corresponding results for the degenerate four-level engine without repeating the general procedure.

\subsection{Three-level model}

\subsubsection{ Current statistics and KUR quantifier}

A single counting field, assigned to either reservoir, suffices for our calculation. Following Refs.~\cite{Bruderer2014,SchallerBook}, we assign the counting field $\chi_c$ to the cold reservoir. 
We define the net photon transfer to the cold reservoir as $N_c=N_c^{\rm em}-N_c^{\rm abs}$, where $N_c^{\rm em}$ and $N_c^{\rm abs}$ denote the numbers of emission and absorption events, respectively, and adopt the generating function $Z(\chi_c,t)=\langle e^{-i\chi_c N_c(t)}\rangle$. Accordingly, emission into the cold reservoir carries a factor $e^{-i\chi_c}$, whereas absorption from the cold reservoir carries
$e^{+i\chi_c}$.
The resulting counting-field-dependent Lindblad master equation for the three-level model reads \cite{VSTUR2023}
\begin{equation}
\dot{\rho}=-i[V_R,\rho]+\mathcal{L}_h[\rho]+\mathcal{L}_c^\chi[\rho],
\end{equation}
where $\mathcal{L}_c^\chi$ denotes the counting-field-dependent dissipator associated with the cold reservoir and is given by
\onecolumngrid
\begin{equation}
\mathcal{L}^{\chi}_{c}[\rho]
=\gamma_c(n_c+1)\left(e^{-i\chi}\sigma_{01}\rho\sigma_{01}^{\dagger}
-\frac12\{\sigma_{01}^{\dagger}\sigma_{01},\rho\}\right)
+\gamma_c n_c\left(e^{i\chi}\sigma_{01}^{\dagger}\rho\sigma_{01}
-\frac12\{\sigma_{01}\sigma_{01}^{\dagger},\rho\}\right).
\label{D2A}
\end{equation}

Writing the density matrix in vectorized form as $\rho=(\rho_{11},\rho_{00},\rho_{gg},\rho_{0g},\rho_{g0})^T$, we recast the Lindblad master equation as a linear matrix equation governed by the Liouvillian supermatrix $\textsf{L}(\chi_c)$ as $\dot \rho = \textsf{L}(\chi_c) \rho$, where
\begin{equation} \label{EqFCSLiouvillian2}
\textsf{L}(\chi_c)=
\begin{pmatrix}
-[\gamma_h(n_h+1)+\gamma_c(n_c+1)]
& \gamma_c n_c e^{i\chi_c}
& \gamma_h n_h
& 0 & 0
\\
\gamma_c(n_c+1)e^{-i\chi_c}
& -\gamma_c n_c
& 0
& i\lambda & -i\lambda
\\
\gamma_h(n_h+1)
& 0
& -\gamma_h n_h
& -i\lambda & i\lambda
\\
0
& i\lambda
& -i\lambda
& -\frac{1}{2}(\gamma_h n_h+\gamma_c n_c)
& 0
\\
0
& -i\lambda
& i\lambda
& 0
& -\frac{1}{2}(\gamma_h n_h+\gamma_c n_c)
\end{pmatrix}.
\end{equation}
For long observation times, the integrated cold-reservoir photon count has the $k$th cumulant \cite{SchallerBook}
\begin{equation}
C_k(t) = \left.(i\partial_{\chi_c})^k \ln Z(\chi_c,t)\right|_{\chi_c=0} \simeq
\left.t(i\partial_{\chi_c})^k\xi(\chi_c)\right|_{\chi_c=0},
\label{cg1}
\end{equation}
where $\xi(\chi_c)$ is the eigenvalue of the tilted Liouvillian $\textsf{L}(\chi_c)$ with the largest real part and serves as the scaled cumulant generating function in the long-time limit, $\lim_{t\to\infty}t^{-1}\ln Z(\chi_c,t)=\xi(\chi_c)$.
The first derivative of $\xi(\chi_c)$ with respect to the counting field, evaluated at $\chi_c=0$, gives the mean current $I$, while the second derivative yields the scaled current variance,
$\Delta I=\lim_{t\to\infty}t\,\langle[I(t)-I]^2\rangle$:
\begin{equation}
I=
\left.i\partial_{\chi_c}\xi(\chi_c)\right|_{\chi_c=0} \equiv \xi^\prime, \qquad \Delta I = \left.(i\partial_{\chi_c})^2\xi(\chi_c)\right|_{\chi_c=0} \equiv \xi''.
\label{EqTURCumulant}
\end{equation}

Following Ref.~\cite{Bruderer2014}, we calculate the mean current and its variance using the characteristic polynomial associated with the tilted Liouvillian $\textsf{L}(\chi_c)$:
\begin{equation}
\sum_n a_n(\chi_c)\xi^n(\chi_c)=0.
\label{sum}
\end{equation}
Here, the coefficients $a_n$ depend on the counting field $\chi_c$. We define their first and second derivatives as
\begin{equation}
a_n'\equiv i\partial_{\chi_c}a_n\Big|_{\chi_c=0},
\qquad
a_n''\equiv (i\partial_{\chi_c})^2a_n\Big|_{\chi_c=0}.
\end{equation}
Taking the first and second derivatives of Eq.~\eqref{sum} with respect to the counting field $\chi_c$ and evaluating them at $\chi_c=0$, we obtain
\begin{equation}
\left[i\partial_{\chi_c}\sum_n a_n\xi^n\right]_{\chi_c=0}
=\sum_n[a_n'+(n+1)a_{n+1}\xi']\xi^n(0)=0,
\label{EqFCSfirst}
\end{equation}
and
\begin{equation}
\left[(i\partial_{\chi_c})^2\sum_n a_n\xi^n\right]_{\chi_c=0}
=\sum_n[a_n''+2(n+1)a_{n+1}'\xi'
+(n+1)a_{n+1}\xi''
+(n+1)(n+2)a_{n+2}\xi'^2]\xi^n(0)=0.
\label{EqFCSsecond}
\end{equation}
Since the steady state corresponds to that with the zero eigenvalue of the untilted Liouvillian, we have $\xi(0)=0$. Therefore, the coefficient of the $\xi^0$ term in Eq.~\eqref{EqFCSfirst} must vanish, which yields the mean current, $I=\xi'$:
\begin{equation}
a_0'+a_1\xi'=0
\quad\Rightarrow\quad
\langle I\rangle=\xi'=-\frac{a_0'}{a_1}.
\label{EqFCSmean}
\end{equation}
Similarly, Eq.~\eqref{EqFCSsecond} yields the scaled current variance:
\begin{equation}
\Delta I=\xi''
=-\frac{a_0''+2I(a_1'+a_2I)}{a_1}.
\label{EqFCSvariance}
\end{equation}

For the resonant three-level engine, the coefficients required to evaluate the mean current and its variance are
\begin{equation}
\begin{aligned}
a_0'&=-\lambda^2\gamma_c\gamma_h(n_h-n_c)B',\\
a_0''&=-\lambda^2\gamma_c\gamma_h B'
\left(2n_cn_h+n_c+n_h\right),\\
a_1&=\frac{B'}{4}
\left(B'C'\gamma_c\gamma_h+4\lambda^2D'\right),\\
a_1'&=-2\lambda^2\gamma_c\gamma_h(n_h-n_c),\\
a_2&=\left(\frac{B'^2}{4}+4\lambda^2\right)
\left(2B'+\gamma_c+\gamma_h\right)
+B'\gamma_c\gamma_h C'.
\end{aligned}
\label{eq:fcs_coefficients_3ls}
\end{equation}
Here, the coefficients and their counting-field derivatives are evaluated at $\chi_c=0$, with $B'=\gamma_cn_c+\gamma_hn_h$, $C'=3n_cn_h+n_c+n_h$, and $D'=\gamma_c(2+3n_c)+\gamma_h(2+3n_h)$. 

Applying this method to the Liouvillian in Eq.~\eqref{EqFCSLiouvillian2}, we obtain closed-form expressions for the mean current and variance:  
\begin{equation}
 I
=\frac{4(n_h-n_c)\gamma_h\gamma_c\lambda^2}
{4\lambda^2[\gamma_h(3n_h+2)+\gamma_c(3n_c+2)]
+(3n_hn_c+n_h+n_c)(\gamma_hn_h+\gamma_cn_c)\gamma_h\gamma_c}
\equiv
\frac{4(n_h-n_c)\gamma_h\gamma_c\lambda^2}
{4\lambda^2D'+B'C'\gamma_h\gamma_c},
\label{CurrentII}
\end{equation}
\begin{equation}
\Delta I=\frac{4\gamma_c\gamma_h\lambda^2}{4\lambda^2D'+B'C'\gamma_c\gamma_h}\left[2n_hn_c+n_h+n_c-\frac{8\lambda^2(n_h-n_c)^2\gamma_c\gamma_hG'}{(4\lambda^2D'+B'C'\gamma_c\gamma_h)^2}\right]. \label{variance}
\end{equation}
Here, $G'=8\lambda^2+\gamma_c\gamma_h(10n_hn_c+3n_c+3n_h)+\gamma_c^2(2n_c+1)n_c+\gamma_h^2(2n_h+1)n_h$. Further, dynamical activity for the three-level engine model is defined as
\begin{equation}
\begin{aligned}
A &= \gamma_h(n_h+1)\operatorname{Tr}[\sigma_{g1}\rho\sigma_{g1}^{\dagger}] + \gamma_h n_h\operatorname{Tr}[\sigma_{g1}^{\dagger}\rho\sigma_{g1}] + \gamma_c(n_c+1)\operatorname{Tr}[\sigma_{01}\rho\sigma_{01}^{\dagger}] + \gamma_c n_c\operatorname{Tr}[\sigma_{01}^{\dagger}\rho\sigma_{01}] \\
&= [\gamma_h(n_h+1)  + \gamma_c(n_c+1)]\rho_{11} +  \gamma_h n_h \rho_{gg} +\gamma_c n_c  \rho_{00}.
\end{aligned}
\end{equation}
Using Eqs. (\ref{rhoggII})--(\ref{rho11II}), the final expression for dynamical activity is obtained as
\begin{equation}
A=
\frac{2(\gamma_cn_c+\gamma_hn_h)
[\gamma_c(1+n_c)+\gamma_h(1+n_h)]
(4\lambda^2+n_cn_h\gamma_c\gamma_h)}
{4\lambda^2[\gamma_h(3n_h+2)+\gamma_c(3n_c+2)]
+(3n_hn_c+n_h+n_c)(\gamma_hn_h+\gamma_cn_c)\gamma_h\gamma_c} \equiv \frac{ 2 B\, B'\, H'}{4\lambda^2D'+B'C'\gamma_h\gamma_c} ,
\label{DAII}
\end{equation}
with   $B=\gamma_c(1+n_c)+\gamma_h(1+n_h)$ and $H'=4\lambda^2+n_cn_h\gamma_c\gamma_h$.

The final expression for KUR quantifier  $\mathcal{Q}=I^2/A\Delta I$ can be obtained by combining Eqs.~\eqref{CurrentII}, \eqref{variance} and \eqref{DAII}, and is given by 
\begin{equation}
\mathcal Q
=\frac{2(n_h-n_c)^2\gamma_c\gamma_h\lambda^2}{BB'H'}
\left[
2n_hn_c+n_h+n_c
-\frac{8\lambda^2(n_h-n_c)^2\gamma_c\gamma_hG'}
{(4\lambda^2D'+B'C'\gamma_h\gamma_c)^2}
\right]^{-1},
\label{KURPatrickA}
\end{equation}
 as in Eq.~\eqref{KURPatrick}.

\subsubsection{ Evaluation of the $\chi$ and $\psi$ corrections} \label{sec:subsub_chi_psi}

 Having obtained the current statistics, we now evaluate the corrections entering the quantum KURs. First, the Hasegawa correction $\chi$ is constructed from the auxiliary superoperators ${\mathcal L}_1$ and ${\mathcal L}_2$, with ${\mathcal L}={\mathcal L}_1+{\mathcal L}_2$, defined by
\begin{eqnarray}
{\mathcal L}_1\rho
&=&-i V_R\rho
+\frac12\sum_{k=c,h}\gamma_k(n_k+1)
\left(L_k\rho L_k^\dagger-L_k^\dagger L_k\rho\right)
+\frac12\sum_{k=c,h}\gamma_kn_k
\left(L_k^\dagger\rho L_k-L_kL_k^\dagger\rho\right),\label{EqL1L1}
\\
{\mathcal L}_2\rho
&=&i \rho V_R
+\frac12\sum_{k=c,h}\gamma_k(n_k+1)
\left(L_k\rho L_k^\dagger-\rho L_k^\dagger L_k\right)
+\frac12\sum_{k=c,h}\gamma_kn_k
\left(L_k^\dagger\rho L_k-\rho L_kL_k^\dagger\right).
\label{EqL1L2}
\end{eqnarray}
Here, $L_h=\sigma_{g1}=|g\rangle\langle 1|$ and
$L_c=\sigma_{01}=|0\rangle\langle 1|$ are the jump operators. The Drazin inverse ${\mathcal L}^{D}$ satisfies ${\mathcal L}{\mathcal L}^{D} ={\mathcal L}^{D}{\mathcal L}=\mathds{1}-\mathcal P$ and ${\mathcal L}^{D}\mathcal P
=\mathcal P {\mathcal L}^{D}=0$, where $\mathcal P\rho=\rho_{\rm ss}\operatorname{Tr}\rho$ projects onto the stationary-state subspace. Equivalently, ${\mathcal L}^{D}$ acts as the inverse of
${\mathcal L}$ on the traceless subspace and annihilates the stationary state $\rho_{\rm ss}$~\cite{LandiReview2023}. 
In practice, the action of the Drazin inverse on an operator $X$ can be obtained without explicitly constructing ${\mathcal L}^{D}$: the operator $Y={\mathcal L}^{D}X$ is determined by solving the linear equations
\begin{equation}
    {\mathcal L}Y=(\mathds{1}-\mathcal P)X,
    \qquad
    \operatorname{Tr}Y=0.
\end{equation}
Accordingly, for the two terms entering Eq.~\eqref{QuantumCoherence}, we set $X_i={\mathcal L}_i\rho_{\rm ss}$ and determine $Y_i={\mathcal L}^{D}X_i$ $(i=1,2)$ from
\begin{equation}
    {\mathcal L}Y_i=(\mathds{1}-\mathcal P)X_i,
    \qquad
    \operatorname{Tr}Y_i=0.
\end{equation}
Using the auxiliary superoperators together with the stationary-state solution in Eqs.~\eqref{cohII}--\eqref{rho11II}, the correction is then evaluated as
\begin{equation}
    \chi=-4\left[
    \operatorname{Tr}\!\left({\mathcal L}_1Y_2\right)
    +
    \operatorname{Tr}\!\left({\mathcal L}_2Y_1\right)
    \right],
\end{equation}
which yields the closed-form expression in Eq.~\eqref{chiclosed}.

To evaluate $\psi$ in Eq.~\eqref{psiterm}, we use the rotating-frame Hamiltonian superoperator $\mathcal H X=-i[V_R,X]$ and introduce $Y_\psi=\mathcal L^D \mathcal H\rho_{\rm ss}$. Since $\mathcal H\rho_{\rm ss}$ is traceless, $Y_\psi$ is obtained
from the linear equations
\begin{equation}
\mathcal L Y_\psi=-i[V_R,\rho_{\rm ss}],
\qquad
\operatorname{Tr}Y_\psi=0.
\label{eq:psi_linear_solve}
\end{equation}
For a unique stationary state, the trace condition removes the freedom to add an arbitrary multiple of $\rho_{\rm ss}$ and thereby selects the solution corresponding to the Drazin inverse. Counting photon emission into the cold reservoir positively and absorption negatively, the current superoperator acts as
\begin{equation}
\mathcal JX
=\gamma_c(n_c+1)\sigma_{01}X\sigma_{10}
-\gamma_cn_c\sigma_{10}X\sigma_{01},
\label{eq:psi_current_superoperator}
\end{equation}
where $\sigma_{ij}=\ket{i}\bra{j}$. Consequently,
\begin{equation}
\psi
=\frac{\operatorname{Tr}(\mathcal JY_\psi)}{I}
=\frac{\gamma_c(n_c+1)(Y_\psi)_{11}
-\gamma_cn_c(Y_\psi)_{00}}{I}.
\label{eq:psi_evaluation}
\end{equation}
Solving Eq.~\eqref{eq:psi_linear_solve} using the stationary state
and the rotating-frame dynamics of the three-level engine then yields
the closed-form expression for $(1+\psi)^2$ in
Eq.~\eqref{psiclosed}.

\subsubsection{Evaluation of the $\delta$ correction}
\label{app:delta_evaluation}

The correction $\delta$ in Eq.~\eqref{deltaterm} can be evaluated
using the stationary quantum-reset construction of
Ref.~\cite{KwonLee2025}. In this section, we present the procedure for the
three-level engine.

Including the corresponding transition rates, the jump operators that resolve the emission and absorption processes are
\begin{equation}
\begin{aligned}
L_{h\downarrow}&=\sqrt{\gamma_h(n_h+1)}\ket{g}\bra{1},
&
L_{h\uparrow}&=\sqrt{\gamma_h n_h}\ket{1}\bra{g},\\
L_{c\downarrow}&=\sqrt{\gamma_c(n_c+1)}\ket{0}\bra{1},
&
L_{c\uparrow}&=\sqrt{\gamma_c n_c}\ket{1}\bra{0}.
\end{aligned}
\label{eq:delta_jumps_3ls}
\end{equation}
A jump through channel $m$ resets the system to a definite post-jump state $\rho_m(0)$, independently of the pre-jump state. The subsequent no-jump evolution therefore starts from this well-defined reset state and is governed by the non-Hermitian effective Hamiltonian 
\begin{equation}
H_{\rm eff}\equiv V_R-\frac{i}{2}K, 
\end{equation}
until the next jump occurs, where
\begin{equation}
K \equiv \sum_k L_k^\dagger L_k
\end{equation}
is the total jump-rate operator. Then, the unnormalized state given that no further jump has occurred up to time $t$ can be written as
\begin{equation}
R_m(t)= e^{-iH_{\rm eff}t}\rho_m(0)e^{iH_{\rm eff}^\dagger t},
\label{eq:unnormalized}
\end{equation}
and its normalized state is given by 
\begin{equation}
\rho_m(t)=\frac{R_m(t)}{S_m(t)},
\label{eq:normalized}
\end{equation}
where $S_m(t) \equiv \operatorname{Tr}R_m(t)$ is the corresponding survival probability.

The measured current counts photon emission into the cold reservoir with weight $+1$ and absorption from it with weight $-1$, while hot-bath jumps carry zero current weight. The corresponding current superoperator is therefore
\begin{equation}
\mathcal J = \mathcal J_{c\downarrow} - \mathcal J_{c\uparrow},
\end{equation}
where the jump superoperator associated with channel $k$ is defined as
\begin{equation}
\mathcal J_k X \equiv L_kXL_k^\dagger.
\end{equation}
The stationary jump rate through channel $m$ is then given by
\begin{equation}
f_m =\operatorname{Tr}(\mathcal J_m\rho_{\rm ss}).
\label{eq:stationary_rate}
\end{equation}

To evaluate the correction $\delta$, we first introduce
\begin{equation}
M(t)=\langle \rho_c(t)Z_t\rangle ,
\end{equation}
where $\rho_c (t)$ denotes the density matrix conditioned on the quantum-jump record, as introduced in Eq.~\eqref{eq:delta_centered_jumps}, and obeys the stochastic master equation
\begin{equation}
d\rho_c = \mathcal L\rho_c\,dt +
\sum_k \left( \frac{\mathcal J_k\rho_c}{r_k} -\rho_c \right) d\widetilde N_k.
\end{equation}
Using the It\^o product rule $dM = \langle Z_t d\rho_c \rangle + \langle  \rho_c dZ_t \rangle + \langle d\rho_c dZ_t  \rangle$, we obtain
\begin{equation}
\dot M = \mathcal L M+\mathcal B_\delta,
\label{eq:M_equation}
\end{equation}
where the source term $\mathcal B_\delta$ is given by
\begin{equation}
\begin{aligned}
\mathcal B_\delta & \equiv
\sum_k \mathcal J_k\rho_{\rm ss} - \mathcal C_\delta,\\
\mathcal C_\delta &\equiv
\sum_m f_m \int_0^\infty dt\, S_m(t) \rho_m(t)
\operatorname{Tr}[K\rho_m(t)] = \left\langle \rho_c\,\operatorname{Tr}(K\rho_c) \right\rangle_{\rm ss}.
\end{aligned}
\label{eq:delta_source}
\end{equation}
Here, we have used the fact that stationary averages over the conditioned states can be evaluated by weighting each no-jump trajectory segment by $f_m S_m(t)$. The quantity $\mathcal C_\delta$ is a conditioned-state correlation and therefore cannot, in general, be factorized as $\rho_{\rm ss}\operatorname{Tr}(K\rho_{\rm ss})$.

In the long time limit, where $\dot M_\infty = 0$, Eq.~\eqref{eq:M_equation} can be manipulated as
\begin{equation}
\mathcal L^D \mathcal L M_\infty = (\mathds{1}-\mathcal P)M_\infty = - \mathcal L^D \mathcal B_\delta,
\end{equation}
where $\mathcal L^D$ denotes the Drazin inverse of the Liouvillian. Since $\operatorname{Tr}M=\langle Z_t \operatorname{Tr}\rho_c(t)\rangle= \langle Z_t\rangle=0$, $\mathcal P M_\infty = \rho_{\rm ss} \operatorname{Tr} M_\infty =0$. Therefore, 
\begin{equation}
M_\infty = -\mathcal L^D\mathcal B_\delta.
\label{eq:M_infty}
\end{equation}

The numerator of the correction $\delta$ can be written as
\begin{equation}
\left\langle J_\tau^{\rm dr}Z_\tau\right\rangle= \int_0^\tau dt\, \left\langle \operatorname{Tr}\!\left[\mathcal J\rho_c(t)\right]Z_\tau \right\rangle,
\end{equation}
since the drift contribution to the integrated current is
\begin{equation}
J_\tau^{\rm dr} = \int_0^\tau dt\, \sum_k w_k r_k(t) = \int_0^\tau dt\, \operatorname{Tr}\!\left[\mathcal J\rho_c(t)\right].
\end{equation}
Using $Z_\tau=Z_t+\left(Z_\tau-Z_t\right)$, the integrand can be decomposed as
\begin{align}
\left\langle \operatorname{Tr}\!\left[\mathcal J\rho_c(t)\right]Z_\tau \right\rangle
= \left\langle \operatorname{Tr}\!\left[\mathcal J\rho_c(t)\right]Z_t \right\rangle + \left\langle \operatorname{Tr}\!\left[\mathcal J\rho_c(t)\right] \left(Z_\tau-Z_t\right) \right\rangle.
\end{align}
Since $Z_t=\sum_k\int_0^t d\widetilde N_k(s)$ is a martingale, its future increment has zero conditional mean, $\mathbb E\!\left[Z_\tau-Z_t\,\middle|\,\mathcal F_t\right]=0$, where $\mathcal F_t$ denotes the quantum-jump record up to time $t$. Because $\rho_c(t)$ is determined by $\mathcal F_t$, it follows that $\left\langle \operatorname{Tr}\!\left[\mathcal J\rho_c(t)\right] \left(Z_\tau-Z_t\right) \right\rangle =0$.
Hence,
\begin{equation}
\left\langle J_\tau^{\rm dr}Z_\tau\right\rangle =
\int_0^\tau dt\, \left\langle \operatorname{Tr}\!\left[\mathcal J\rho_c(t)\right]Z_t
\right\rangle = \int_0^\tau dt\,  \operatorname{Tr}\!\left[\mathcal J M(t)\right].
\end{equation}
In the long time limit, 
\begin{equation}
\int_0^\tau dt\, \operatorname{Tr}\!\left[\mathcal J M(t)\right] = \tau\,\operatorname{Tr}\!\left[\mathcal J M_\infty\right] +O(1),
\end{equation}
therefore, we have 
\begin{equation}
\lim_{\tau\to\infty} \frac{ \left\langle J_\tau^{\rm dr}Z_\tau\right\rangle }{\tau}
= \operatorname{Tr}\!\left[\mathcal J M_\infty\right].
\end{equation}
Using Eq.~\eqref{eq:M_infty}, the correction becomes
\begin{equation}
\delta = -\frac{ \operatorname{Tr} \left[ \mathcal J\mathcal L^D\mathcal B_\delta \right] }{I},
\qquad I=\operatorname{Tr}(\mathcal J\rho_{\rm ss}).
\label{eq:delta_computational}
\end{equation}
In practice, the Drazin inverse need not be constructed explicitly. Instead, one may determine $Y=\mathcal L^D \mathcal B_\delta$ by solving $\mathcal L Y= \mathcal B_\delta$ and $\operatorname{Tr}Y=0$, and then evaluate
$\delta = -\operatorname{Tr}[\mathcal JY]/I$.

The above procedure applies in the relaxing regime, where the system has a unique stationary state and the no-jump survival probabilities decay at long times. Its numerical implementation requires the stationary state, the time integral in Eq.~\eqref{eq:delta_source}, and a single traceless linear solve. Numerical convergence can be checked using
\begin{equation}
\sum_m f_m\int_0^\infty S_m(t)\,dt=1, 
\qquad \sum_m f_m\int_0^\infty R_m(t)\,dt=\rho_{\rm ss}, \qquad \operatorname{Tr}\mathcal B_\delta=0.
\end{equation}

\subsection{Four-level degenerate engine}

The current and its variance can be evaluated using the same full counting statistics as for the three-level engine. A counting field is attached to the cold-bath jump operator. Since the cold bath still mediates only the $\ket{1}\leftrightarrow\ket{0}$ transition, $\mathcal L_c^\chi$ is identical to Eq.~\eqref{D2A}. The mean current and its variance then follow from the eigenvalue of the tilted Liouvillian with the largest real part, exactly as in Eqs.~\eqref{cg1}--\eqref{EqFCSvariance}. The larger Hilbert space, however---involving the populations $\rho_{11},\rho_{00},\rho_{g_1g_1},\rho_{g_2g_2}$, together with the noise-induced coherence $\rho_{g_1g_2}$ and the lasing coherences $\rho_{0g_1},\rho_{0g_2}$---makes the tilted Liouvillian and its characteristic polynomial considerably more cumbersome than the $5\times5$ matrix in Eq.~\eqref{EqFCSLiouvillian2}. We therefore omit their explicit forms and present only the final closed-form results.

It is convenient to introduce $X\equiv\gamma_cn_c+\gamma_hn_h(1+p)$, $W\equiv\gamma_c(4n_cn_h+n_c+2n_h)+\gamma_hn_h(4n_h+3)(1+p)$, and the common denominator $\Xi\equiv\gamma_c\gamma_hn_h(1+p)X(4n_cn_h+2n_c+n_h)+8\lambda^2W$.
In terms of these quantities, the current and dynamical activity are given by
\begin{equation}
 I
=\frac{8\gamma_c\gamma_h\lambda^2n_h(n_h-n_c)(1+p)}{\Xi},
\label{eq:Ideg}
\end{equation}
and
\begin{equation}
A =
\frac{2n_hX[\gamma_c(1+n_c)+2\gamma_h(1+n_h)]
[\gamma_c\gamma_hn_cn_h(1+p)+8\lambda^2]}{\Xi}.
\label{eq:Adeg}
\end{equation}
The Fano factor takes the form
\begin{equation}
F =\frac{2n_cn_h+n_c+n_h}{n_h-n_c}
+\frac{\lambda^2(\mathcal G_0+\lambda^2\mathcal G_2)}
{(n_h-n_c)(1-p)\Xi^2},
\label{eq:Fdeg}
\end{equation}
where
\begin{eqnarray}
\mathcal G_2
&\equiv&
128\gamma_c(n_h-n_c)^2(1+p)
\left[\gamma_cn_c(n_h+1)
+\gamma_hn_h\bigl(p(5n_h+3)-(3n_h+1)\bigr)\right],
\label{eq:G2def}
\\
\mathcal G_0
&\equiv&
16\gamma_c\gamma_hn_h(n_h-n_c)^2(1+p)K_0,
\qquad
K_0\equiv
\gamma_c^2n_c\left[2n_c\bigl(p(2n_h+1)-n_h\bigr)+n_h(p-1)\right]
\nonumber\\
&&\quad
+\gamma_c\gamma_hn_h(1+p)
\left[n_c\bigl(p(15n_h+8)-(13n_h+6)\bigr)+3n_h(p-1)\right]
+\gamma_h^2n_h^2(3n_h+2)(p-1)(1+p)^2.
\label{eq:G0def}
\end{eqnarray}
As in the three-level case, the   KUR quantifier is $\mathcal Q=I/(AF)$. Since $I/A\leq 1$, a violation of the standard KUR, $\mathcal Q>1$, necessarily requires $F<1$. Combining Eqs.~\eqref{eq:Ideg}--\eqref{eq:Fdeg}, we obtain
the inverse KUR quantifier for the degenerate four-level engine:
\begin{equation}
\mathcal Q
= \frac{I^2}{A\Delta I}
=\frac{I}{AF}
=
\frac{4\gamma_c\gamma_h\lambda^2(n_h-n_c)^2(1+p)}
{X\bigl[\gamma_c(1+n_c)+2\gamma_h(1+n_h)\bigr]
\bigl[\gamma_c\gamma_h n_c n_h(1+p)+8\lambda^2\bigr]}
\left[
2n_cn_h+n_c+n_h
+\frac{\lambda^2\bigl(\mathcal G_0+\lambda^2\mathcal G_2\bigr)}
{(1-p)\Xi^2}
\right]^{-1}.
\label{eq:Qdeg_explicit}
\end{equation}

The corrections ${\chi}$ and $\psi$ for the four-level engine are obtained by following the same procedure as for the three-level engine. Since the cold reservoir couples the same pair of levels, the current superoperator in Eq.~\eqref{eq:psi_current_superoperator} remains unchanged. The coherence-dependent terms entering the Hasegawa and $\psi$-KUR bounds are then given by
\begin{equation}
{\chi}
=\frac{64 n_h\lambda^2\,Z[8\lambda^2 + \gamma_c\gamma_h n_c n_h (1+p)]}
{X\Xi},
\label{eq:chideg}
\end{equation}
with   $Z\equiv\gamma_c(1+n_c)+\gamma_h(1+n_h)(1+p)$, and
\begin{equation}
(1+\psi)^2
=\left(\frac{\Xi-16\lambda^2W}{\Xi}\right)^2.
\label{eq:psideg}
\end{equation}

For the $\delta$ correction, the procedure in Appendix~\ref{app:delta_evaluation} applies after replacing the three-level hot-reservoir jumps by the resolved four-level channels. For equal hot-reservoir couplings, diagonalizing the hot-reservoir rate matrix $\gamma_h\bigl(\begin{smallmatrix}1&p\\p&1\end{smallmatrix}\bigr)$ gives
\begin{equation}
\begin{aligned}
\ket{g_\pm}
&=\frac{\ket{g_1}\pm\ket{g_2}}{\sqrt2},\\
L_{h,\pm\downarrow}
&=\sqrt{\gamma_h(1\pm p)(n_h+1)}\ket{g_\pm}\bra{1},\\
L_{h,\pm\uparrow}
&=\sqrt{\gamma_h(1\pm p)n_h}\ket{1}\bra{g_\pm}.
\end{aligned}
\label{eq:delta_jumps_4ls}
\end{equation}
The cold-reservoir jump operators remain those in Eq.~\eqref{eq:delta_jumps_3ls}. These resolved channels also reset the system to definite post-jump states. Using these jump operators together with the four-level Hamiltonian and following the same procedures as in Appendix~\ref{app:delta_evaluation} determines $\delta$ without repeating the derivation.

\section{Proof of the tighter $\psi$-KUR}
\label{app:tighter_psi_kur}

We prove the $\psi$-bound in Eq.~\eqref{eq:tighter_psi_kur} for
$n_h>n_c>0$ and $\gamma_c,\gamma_h,\lambda>0$.
For notational convenience, we introduce
\begin{equation}
\begin{aligned}
z&=4\lambda^2,
& q&=n_cn_h,
& g&=\gamma_c\gamma_h,&
d_n&=n_h-n_c,
& u&=\gamma_c+\gamma_h,
\end{aligned}
\label{eq:psi_auxiliary_a}
\end{equation}
together with
\begin{equation}
\begin{aligned}
s&=n_c\gamma_c+n_h\gamma_h,&
B&=u+s,
& C&=2u+3s,\\
T&=n_c+n_h+2q,
& E&=n_c+n_h+3q,&
P&=Egs.
\end{aligned}
\label{eq:psi_auxiliary_b}
\end{equation}
All quantities defined above are strictly
positive under the stated assumptions.
The explicit expression for $\mathcal{Q}_\psi$ can be written in the compact form
\begin{equation}
\mathcal{Q}_{\psi}
=
\frac{d_n^2gz(P-Cz)^2}
{2Bs(q g+z)\mathscr{D}},
\label{eq:psi_compact}
\end{equation}
where
\begin{equation}
\begin{aligned}
\mathscr{D}
={}&T(P+Cz)^2 -2d_n^2gz
\left[s(u+2s)+2Eg+2z\right].
\end{aligned}
\label{eq:psi_denominator}
\end{equation}

Because $\mathscr{D}$ is expressed as the difference of two positive terms, we first establish its positivity before deriving an upper bound on $\mathcal{Q}_\psi$. A convenient way to do this is to compare $Bs\mathscr{D}$ with the numerator factor $d_n^2g(P-Cz)^2$ appearing in Eq.~\eqref{eq:psi_compact}. We therefore define
\begin{equation}
\mathscr{W}
\equiv
Bs\mathscr{D}-d_n^2g(P-Cz)^2.
\label{eq:psi_positive_difference}
\end{equation}
Expanding Eq.~\eqref{eq:psi_positive_difference} in powers of $z$ and collecting terms in ascending powers of $z$ yields
\begin{equation}
\mathscr{W}=a_0P^2+2a_1z+a_2z^2,
\label{eq:psi_positive_polynomial}
\end{equation}
where
\begin{subequations}
\label{eq:psi_positive_coefficients}
\begin{align}
a_0={}&(T+1)s^2+q u(u+2s),
\\
a_1={}&gs^2\Big[
TEB^2+(TE-d_n^2)B(u+2s)+d_n^2Eg
\Big],
\\
a_2={}&\left[s^2+q u(u+2s)\right]
(C^2+4Bs) +Ts^3(8B+s).
\end{align}
\end{subequations}
The coefficients $a_0$ and $a_2$ are manifestly positive. The coefficient $a_1$ is also positive, since 
\begin{equation}
TE-d_n^2
= q\left[4+5(n_c+n_h)+6 q\right]>0.
\label{eq:psi_TE_positive}
\end{equation}
Therefore, $\mathscr{W}>0$, which implies 
\begin{equation}
Bs\mathscr{D}
> d_n^2g(P-Cz)^2
\geq0.
\label{eq:psi_denominator_comparison}
\end{equation}
Since $B>0$ and $s>0$, we conclude that $\mathscr{D}>0$.
Equation~\eqref{eq:psi_compact} then gives
\begin{equation}
\begin{aligned}
0\leq\mathcal{Q}_{\psi}
&=
\frac{z}{2(qg+z)}
\frac{d_n^2g(P-Cz)^2}{Bs\mathscr{D}}<
\frac{z}{2(qg+z)}
<
\frac12.
\end{aligned}
\label{eq:psi_bound_chain}
\end{equation}
Restoring the original parameters, we obtain the stronger parameter-dependent bound
\begin{equation}
\mathcal{Q}_{\psi}
<
\frac{2\lambda^2}
{n_cn_h\gamma_c\gamma_h+4\lambda^2}
<
\frac12,
\label{eq:psi_parameter_bound}
\end{equation}
which establishes the $\psi$-bound in Eq.~\eqref{eq:tighter_psi_kur} throughout the stated parameter domain.

Finally, we establish the sharpness of the constant $1/2$. For the parameter family in Eq.~\eqref{eq:psi_sharp_family}, $z=4\Gamma^2$ and, as $\varepsilon\to0^+$,
\begin{equation}
\begin{aligned}
s&\sim\varepsilon\Gamma,
& B&\to\Gamma,&
g&=\varepsilon\Gamma^2,
& \revblue{d_n}&\to1,\\
P&\sim\varepsilon^2\Gamma^3,
& C&\to2\Gamma,&
\revblue{q}g&=\varepsilon^3\Gamma^2,
& \mathscr{D}&\to64\Gamma^6.
\end{aligned}
\end{equation}
Substituting these asymptotic forms into Eq.~\eqref{eq:psi_compact} yields
\begin{equation}
\lim_{\varepsilon\to0^+}\mathcal{Q}_{\psi}
=\frac12.
\end{equation}
Hence, $1/2$ is the supremum over the stated parameter domain, although it is not attained for any finite set of strictly positive parameters.


 \bibliography{TUR_NIC}

@article{VSTUR2023,
  title = {Thermodynamic uncertainty relation in nondegenerate and degenerate maser heat engines},
  author = {Singh, Varinder and Shaghaghi, Vahid and M\"ustecapl\ifmmode \imath \else \i \fi{}o\ifmmode \breve{g}\else \u{g}\fi{}lu, \"Ozg\"ur E. and Rosa, Dario},
  journal = {Phys. Rev. A},
  volume = {108},
  issue = {3},
  pages = {032203},
  numpages = {12},
  year = {2023},
  month = {Sep},
  publisher = {American Physical Society},
  doi = {10.1103/PhysRevA.108.032203},
  url = {https://link.aps.org/doi/10.1103/PhysRevA.108.032203}
}

@article{KwonLee2025,
  title = {A unified framework for classical and quantum uncertainty relations using stochastic representations},
  author = {Kwon, Euijoon and Lee, Jae Sung},
  journal = {Communications Physics},
  volume = {8},
  pages = {444},
  year = {2025},
  publisher = {Nature Portfolio},
  doi = {10.1038/s42005-025-02348-y},
  url = {https://doi.org/10.1038/s42005-025-02348-y}
}

@article{Skrzypczyk2014,
    author = {Skrzypczyk, Paul and Short, Anthony J. and Popescu, Sandu},
    journal = {Nat. Commun.},
    volume = {5},
    pages = {4185},
    title = {Work extraction and thermodynamics for individual quantum systems},
    year = {2014},
    doi = {10.1038/ncomms5185},
    url = {https://doi.org/10.1038/ncomms5185}
}

@article{Latune2020,
    author = {Latune, C. L. and Sinayskiy, I. and Petruccione, F.},
    journal = {Eur. Phys. J. Spec. Top.},
    volume = {230},
    pages = {841--850},
    title = {Roles of quantum coherences in thermal machines},
    year = {2021},
    doi = {10.1140/epjs/s11734-021-00085-1},
    url = {https://doi.org/10.1140/epjs/s11734-021-00085-1}
}

@article{Lutz2014,
	author = {Ro\ss{}nagel, J. and Abah, O. and Schmidt-Kaler, F. and Singer, K. and Lutz, E.},
	doi = {10.1103/PhysRevLett.112.030602},
	issue = {3},
	journal = {Phys. Rev. Lett.},
	month = {Jan},
	numpages = {5},
	pages = {030602},
	publisher = {American Physical Society},
	title = {Nanoscale Heat Engine Beyond the Carnot Limit},
	url = {https://link.aps.org/doi/10.1103/PhysRevLett.112.030602},
	volume = {112},
	year = {2014}}

@article{Manzano2016,
	author = {Manzano, Gonzalo and Galve, Fernando and Zambrini, Roberta and Parrondo, Juan MR},
	journal = {Phys. Rev. E},
	pages = {052120},
	title = {Entropy production and thermodynamic power of the squeezed thermal reservoir},
	volume = {93},
	year = {2016}}

@article{VOzgur2020,
  title = {Performance bounds of nonadiabatic quantum harmonic Otto engine and refrigerator under a squeezed thermal reservoir},
  author = {Singh, Varinder and M{\"u}stecapl{\ifmmode \imath \else \i \fi{}}o{\ifmmode \breve{g}\else \u{g}\fi{}}lu, {\"O}zg{\"u}r E.},
  journal = {Phys. Rev. E},
  volume = {102},
  issue = {6},
  pages = {062123},
  numpages = {8},
  year = {2020},
  month = {Dec},
  publisher = {American Physical Society},
  doi = {10.1103/PhysRevE.102.062123},
  url = {https://link.aps.org/doi/10.1103/PhysRevE.102.062123}
}

@article{Alonso2014,
    author = {Correa, Luis A. and Palao, Jos{\'e} P. and Alonso, Daniel and Adesso, Gerardo},
    journal = {Sci. Rep.},
    volume = {4},
    pages = {3949},
    title = {Quantum-enhanced absorption refrigerators},
    year = {2014},
    doi = {10.1038/srep03949},
    url = {https://doi.org/10.1038/srep03949}
}

@article{Klaers2017,
    author = {Klaers, Jan and Faelt, Stefan and Imamoglu, Atac and Togan, Emre},
    journal = {Phys. Rev. X},
    volume = {7},
    pages = {031044},
    title = {Squeezed thermal reservoirs as a resource for a nanomechanical engine beyond the {Carnot} limit},
    year = {2017},
    doi = {10.1103/PhysRevX.7.031044},
    url = {https://doi.org/10.1103/PhysRevX.7.031044}
}

@article{OzgurQuantumFuel,
	author = {T\"urkpen\ifmmode \mbox{\c{c}}\else \c{c}\fi{}e, Deniz and M\"ustecapl\ifmmode \imath \else \i \fi{}o\ifmmode \breve{g}\else \u{g}\fi{}lu, \"Ozg\"ur E.},
	doi = {10.1103/PhysRevE.93.012145},
	issue = {1},
	journal = {Phys. Rev. E},
	month = {Jan},
	numpages = {14},
	pages = {012145},
	publisher = {American Physical Society},
	title = {Quantum fuel with multilevel atomic coherence for ultrahigh specific work in a photonic Carnot engine},
	url = {https://link.aps.org/doi/10.1103/PhysRevE.93.012145},
	volume = {93},
	year = {2016}}

@article{LandiReview2023,
  title = {Current Fluctuations in Open Quantum Systems: Bridging the Gap Between Quantum Continuous Measurements and Full Counting Statistics},
  author = {Landi, Gabriel T. and Kewming, Michael J. and Mitchison, Mark T. and Potts, Patrick P.},
  journal = {PRX Quantum},
  volume = {5},
  issue = {2},
  pages = {020201},
  numpages = {86},
  year = {2024},
  month = {Apr},
  publisher = {American Physical Society},
  doi = {10.1103/PRXQuantum.5.020201},
  url = {https://link.aps.org/doi/10.1103/PRXQuantum.5.020201}
}

@article{Vu2025PRX,
  title = {Fundamental Bounds on Precision and Response for Quantum Trajectory Observables},
  author = {Van Vu, Tan},
  journal = {PRX Quantum},
  volume = {6},
  issue = {1},
  pages = {010343},
  numpages = {22},
  year = {2025},
  month = {Mar},
  publisher = {American Physical Society},
  doi = {10.1103/PRXQuantum.6.010343},
  url = {https://link.aps.org/doi/10.1103/PRXQuantum.6.010343}
}

@article{Moreira2025,
  title = {Precision bounds for multiple currents in open quantum systems},
  author = {Moreira, Saulo V. and Radaelli, Marco and Candeloro, Alessandro and Binder, Felix C. and Mitchison, Mark T.},
  journal = {Phys. Rev. E},
  volume = {111},
  issue = {6},
  pages = {064107},
  numpages = {15},
  year = {2025},
  month = {Jun},
  publisher = {American Physical Society},
  doi = {10.1103/PhysRevE.111.064107},
  url = {https://link.aps.org/doi/10.1103/PhysRevE.111.064107}
}

@article{Asli2020,
    author = {\"Ozdemir, A. T. and M\"ustecapl{\i}o\u{g}lu, \"O. E.},
    doi = {10.3906/fiz-2009-12},
    journal = {Turk. J. Phys.},
    number = {5},
    pages = {404--436},
    title = {Quantum thermodynamics and quantum coherence engines},
    url = {https://doi.org/10.3906/fiz-2009-12},
    volume = {44},
    year = {2020}
}

@article{PietzonkaSeifert2016,
	author = {Pietzonka, Patrick and Seifert, Udo},
	doi = {10.1103/PhysRevLett.120.190602},
	issue = {19},
	journal = {Phys. Rev. Lett.},
	month = {May},
	numpages = {6},
	pages = {190602},
	publisher = {American Physical Society},
	title = {Universal Trade-Off between Power, Efficiency, and Constancy in Steady-State Heat Engines},
	url = {https://link.aps.org/doi/10.1103/PhysRevLett.120.190602},
	volume = {120},
	year = {2018}}

@article{Holubec2019,
	author = {Holubec, Viktor and Novotný, Tomáš},
	journal = {J. Chem. Phys.},
	number = {4},
	pages = {044108},
	publisher = {AIP Publishing LLC},
	title = {Effects of noise-induced coherence on the fluctuations of current in quantum absorption refrigerators},
	volume = {151},
	year = {2019}}

@article{Bruderer2014,
    author = {Bruderer, Martin and Contreras-Pulido, L. Debora and Thaller, Maximilian and Sironi, Lucia and Obreschkow, Danail and Plenio, Martin B.},
    journal = {New J. Phys.},
    number = {3},
    pages = {033030},
    title = {Inverse counting statistics for stochastic and open quantum systems: the characteristic polynomial approach},
    volume = {16},
    year = {2014},
    doi = {10.1088/1367-2630/16/3/033030},
    url = {https://doi.org/10.1088/1367-2630/16/3/033030}
}

@book{SchallerBook,
	author = {Schaller, Gernot},
	publisher = {Springer},
	title = {Open quantum systems far from equilibrium},
	volume = {881},
	year = {2014}}

@article{HiuraSasa,
  title = {Kinetic uncertainty relation on first-passage time for accumulated current},
  author = {Hiura, Ken and Sasa, Shin-ichi},
  journal = {Phys. Rev. E},
  volume = {103},
  issue = {5},
  pages = {L050103},
  numpages = {5},
  year = {2021},
  month = {May},
  publisher = {American Physical Society},
  doi = {10.1103/PhysRevE.103.L050103},
  url = {https://link.aps.org/doi/10.1103/PhysRevE.103.L050103}
}

@article{Maes2020,
title = {Frenesy: Time-symmetric dynamical activity in nonequilibria},
journal = {Physics Reports},
volume = {850},
pages = {1-33},
year = {2020},
note = {Frenesy: time-symmetric dynamical activity in nonequilibria},
issn = {0370-1573},
doi = {https://doi.org/10.1016/j.physrep.2020.01.002},
url = {https://www.sciencedirect.com/science/article/pii/S0370157320300120},
author = {Christian Maes}
}

@article{Hasegawa2020,
	author = {Hasegawa, Yoshihiko},
	doi = {10.1103/PhysRevLett.125.050601},
	issue = {5},
	journal = {Phys. Rev. Lett.},
	month = {Jul},
	numpages = {6},
	pages = {050601},
	publisher = {American Physical Society},
	title = {Quantum Thermodynamic Uncertainty Relation for Continuous Measurement},
	url = {https://link.aps.org/doi/10.1103/PhysRevLett.125.050601},
	volume = {125},
	year = {2020}}

@article{Varinder2020,
	author = {Singh, Varinder},
	doi = {10.1103/PhysRevResearch.2.043187},
	issue = {4},
	journal = {Phys. Rev. Research},
	month = {Nov},
	numpages = {12},
	pages = {043187},
	publisher = {American Physical Society},
	title = {Optimal operation of a three-level quantum heat engine and universal nature of efficiency},
	url = {https://link.aps.org/doi/10.1103/PhysRevResearch.2.043187},
	volume = {2},
	year = {2020}}

@article{Davinder2021,
	author = {Singh, Davinder and Hyeon, Changbong},
	doi = {10.1103/PhysRevE.104.054115},
	issue = {5},
	journal = {Phys. Rev. E},
	month = {Nov},
	numpages = {10},
	pages = {054115},
	publisher = {American Physical Society},
	title = {Origin of loose bound of the thermodynamic uncertainty relation in a dissipative two-level quantum system},
	url = {https://link.aps.org/doi/10.1103/PhysRevE.104.054115},
	volume = {104},
	year = {2021}}

@article{FeiGu2025,
  title = {Stochastic Scovil--Schulz-DuBois machine and its three types of cycles},
  author = {Liu, Fei and Gu, Jiayin},
  journal = {Phys. Rev. E},
  volume = {111},
  issue = {1},
  pages = {014108},
  numpages = {9},
  year = {2025},
  month = {Jan},
  publisher = {American Physical Society},
  doi = {10.1103/PhysRevE.111.014108},
  url = {https://link.aps.org/doi/10.1103/PhysRevE.111.014108}
}

@article{BrijMohan2025,
  title = {Coherent heat transfer leads to genuine quantum enhancement in the performances of continuous engines},
  author = {Mohan, Brij and Gangwar, Rajeev and Pandit, Tanmoy and Bera, Mohit Lal and Lewenstein, Maciej and Bera, Manabendra Nath},
  journal = {Phys. Rev. Appl.},
  volume = {23},
  issue = {4},
  pages = {044050},
  numpages = {25},
  year = {2025},
  month = {Apr},
  publisher = {American Physical Society},
  doi = {10.1103/PhysRevApplied.23.044050},
  url = {https://link.aps.org/doi/10.1103/PhysRevApplied.23.044050}
}

@article{Mark2021,
	author = {Rignon-Bret, Antoine and Guarnieri, Giacomo and Goold, John and Mitchison, Mark T.},
	doi = {10.1103/PhysRevE.103.012133},
	issue = {1},
	journal = {Phys. Rev. E},
	month = {Jan},
	numpages = {12},
	pages = {012133},
	publisher = {American Physical Society},
	title = {Thermodynamics of precision in quantum nanomachines},
	url = {https://link.aps.org/doi/10.1103/PhysRevE.103.012133},
	volume = {103},
	year = {2021}}

@article{SinghMilburn2026,
    author = {Singh, Varinder and Kwon, Euijoon and Milburn, G. J.},
    title = {Quantum thermodynamics on a limit cycle},
    journal = {Quantum Sci. Technol.},
    volume = {11},
    number = {2},
    pages = {025026},
    year = {2026},
    month = {Mar},
    publisher = {IOP Publishing},
    doi = {10.1088/2058-9565/ae34df},
    url = {https://doi.org/10.1088/2058-9565/ae34df}
}

@article{HasegawaVu2019,
  title = {Uncertainty relations in stochastic processes: An information inequality approach},
  author = {Hasegawa, Yoshihiko and Van Vu, Tan},
  journal = {Phys. Rev. E},
  volume = {99},
  issue = {6},
  pages = {062126},
  numpages = {12},
  year = {2019},
  month = {Jun},
  publisher = {American Physical Society},
  doi = {10.1103/PhysRevE.99.062126},
  url = {https://link.aps.org/doi/10.1103/PhysRevE.99.062126}
}

@article{JongMin2021,
	author = {Park, Jong-Min and Park, Hyunggyu},
	doi = {10.1103/PhysRevResearch.3.043005},
	issue = {4},
	journal = {Phys. Rev. Research},
	month = {Oct},
	numpages = {15},
	pages = {043005},
	publisher = {American Physical Society},
	title = {Thermodynamic uncertainty relation in the overdamped limit with a magnetic Lorentz force},
	url = {https://link.aps.org/doi/10.1103/PhysRevResearch.3.043005},
	volume = {3},
	year = {2021}}

@article{Garrahan2017,
	author = {Garrahan, Juan P.},
	doi = {10.1103/PhysRevE.95.032134},
	issue = {3},
	journal = {Phys. Rev. E},
	month = {Mar},
	numpages = {7},
	pages = {032134},
	publisher = {American Physical Society},
	title = {Simple bounds on fluctuations and uncertainty relations for first-passage times of counting observables},
	url = {https://link.aps.org/doi/10.1103/PhysRevE.95.032134},
	volume = {95},
	year = {2017}}

@article{Prech2025,
  title = {Role of Quantum Coherence in Kinetic Uncertainty Relations},
  author = {Prech, Kacper and Potts, Patrick P. and Landi, Gabriel T.},
  journal = {Phys. Rev. Lett.},
  volume = {134},
  issue = {2},
  pages = {020401},
  numpages = {8},
  year = {2025},
  month = {Jan},
  publisher = {American Physical Society},
  doi = {10.1103/PhysRevLett.134.020401},
  url = {https://link.aps.org/doi/10.1103/PhysRevLett.134.020401}
}

@article{ScullyAgarwal2003,
  author  = {Scully, Marlan O. and Zubairy, M. Suhail and Agarwal, Girish S. and Walther, Herbert},
  title   = {Extracting Work from a Single Heat Bath via Vanishing Quantum Coherence},
  journal = {Science},
  volume  = {299},
  number  = {5608},
  pages   = {862--864},
  year    = {2003},
  doi     = {10.1126/science.1078955}
}

@article{Shiraishi2016,
  title = {Universal Trade-Off Relation between Power and Efficiency for Heat Engines},
  author = {Shiraishi, Naoto and Saito, Keiji and Tasaki, Hal},
  journal = {Phys. Rev. Lett.},
  volume = {117},
  issue = {19},
  pages = {190601},
  numpages = {6},
  year = {2016},
  month = {Oct},
  publisher = {American Physical Society},
  doi = {10.1103/PhysRevLett.117.190601},
  url = {https://link.aps.org/doi/10.1103/PhysRevLett.117.190601}
}

@article{VuSaito2022,
	author = {Van Vu, Tan and Saito, Keiji},
	doi = {10.1103/PhysRevLett.128.140602},
	issue = {14},
	journal = {Phys. Rev. Lett.},
	month = {Apr},
	numpages = {8},
	pages = {140602},
	publisher = {American Physical Society},
	title = {Thermodynamics of Precision in Markovian Open Quantum Dynamics},
	url = {https://link.aps.org/doi/10.1103/PhysRevLett.128.140602},
	volume = {128},
	year = {2022}}

@article{Patrick2021,
	author = {Kalaee, Alex Arash Sand and Wacker, Andreas and Potts, Patrick P.},
	doi = {10.1103/PhysRevE.104.L012103},
	issue = {1},
	journal = {Phys. Rev. E},
	month = {Jul},
	numpages = {6},
	pages = {L012103},
	publisher = {American Physical Society},
	title = {Violating the thermodynamic uncertainty relation in the three-level maser},
	url = {https://link.aps.org/doi/10.1103/PhysRevE.104.L012103},
	volume = {104},
	year = {2021}}

@article{Horowitz2020,
    author = {Horowitz, Jordan M. and Gingrich, Todd R.},
    journal = {Nat. Phys.},
    volume = {16},
    number = {1},
    pages = {15--20},
    title = {Thermodynamic uncertainty relations constrain non-equilibrium fluctuations},
    year = {2020},
    doi = {10.1038/s41567-019-0702-6},
    url = {https://doi.org/10.1038/s41567-019-0702-6}
}

@article{Gingrich2016,
	author = {Gingrich, Todd R. and Horowitz, Jordan M. and Perunov, Nikolay and England, Jeremy L.},
	doi = {10.1103/PhysRevLett.116.120601},
	issue = {12},
	journal = {Phys. Rev. Lett.},
	month = {Mar},
	numpages = {5},
	pages = {120601},
	publisher = {American Physical Society},
	title = {Dissipation Bounds All Steady-State Current Fluctuations},
	url = {https://link.aps.org/doi/10.1103/PhysRevLett.116.120601},
	volume = {116},
	year = {2016}}

@article{SeifertTUR,
	author = {Barato, Andre C. and Seifert, Udo},
	doi = {10.1103/PhysRevLett.114.158101},
	issue = {15},
	journal = {Phys. Rev. Lett.},
	month = {Apr},
	numpages = {5},
	pages = {158101},
	publisher = {American Physical Society},
	title = {Thermodynamic Uncertainty Relation for Biomolecular Processes},
	url = {https://link.aps.org/doi/10.1103/PhysRevLett.114.158101},
	volume = {114},
	year = {2015}}

@article{Kosloff2013,
    author = {Kosloff, Ronnie},
    title = {Quantum Thermodynamics: A Dynamical Viewpoint},
    journal = {Entropy},
    volume = {15},
    number = {6},
    pages = {2100--2128},
    year = {2013},
    doi = {10.3390/e15062100},
    url = {https://www.mdpi.com/1099-4300/15/6/2100}
}

@article{Scovil1959,
	author = {Scovil, H. E. D. and Schulz-DuBois, E. O.},
	doi = {10.1103/PhysRevLett.2.262},
	issue = {6},
	journal = {Phys. Rev. Lett.},
	month = {Mar},
	numpages = {0},
	pages = {262-263},
	publisher = {American Physical Society},
	title = {Three-Level Masers as Heat Engines},
	url = {https://link.aps.org/doi/10.1103/PhysRevLett.2.262},
	volume = {2},
	year = {1959}}

@article{BoukobzaTannor2007,
	author = {Boukobza, E. and Tannor, D. J.},
	doi = {10.1103/PhysRevLett.98.240601},
	issue = {24},
	journal = {Phys. Rev. Lett.},
	month = {Jun},
	numpages = {4},
	pages = {240601},
	publisher = {American Physical Society},
	title = {Three-Level Systems as Amplifiers and Attenuators: A Thermodynamic Analysis},
	url = {https://link.aps.org/doi/10.1103/PhysRevLett.98.240601},
	volume = {98},
	year = {2007}}

@article{Scully2011,
	author = {Scully, Marlan O. and Chapin, Kimberly R. and Dorfman, Konstantin E. and Kim, Moochan Barnabas and Svidzinsky, Anatoly},
	doi = {10.1073/pnas.1110234108},
	journal = {Proc. Natl. Acad. Sci. USA},
	number = {37},
	pages = {15097-15100},
	publisher = {National Academy of Sciences},
	title = {Quantum heat engine power can be increased by noise-induced coherence},
	volume = {108},
	year = {2011}}

@article{Scully2001,
	author = {Scully, Marlan O.},
	doi = {10.1103/PhysRevLett.87.220601},
	issue = {22},
	journal = {Phys. Rev. Lett.},
	month = {Nov},
	numpages = {4},
	pages = {220601},
	publisher = {American Physical Society},
	title = {Extracting Work from a Single Thermal Bath via Quantum Negentropy},
	url = {https://link.aps.org/doi/10.1103/PhysRevLett.87.220601},
	volume = {87},
	year = {2001}}

@article{Geva1996,
	author = {Geva,Eitan and Kosloff,Ronnie},
	doi = {10.1063/1.471453},
	journal = {J. Chem. Phys.},
	number = {19},
	pages = {7681-7699},
	title = {The quantum heat engine and heat pump: An irreversible thermodynamic analysis of the three‐level amplifier},
	volume = {104},
	year = {1996}}

@article{Geva1994,
	author = {Geva, Eitan and Kosloff, Ronnie},
	doi = {10.1103/PhysRevE.49.3903},
	issue = {5},
	journal = {Phys. Rev. E},
	month = {May},
	numpages = {0},
	pages = {3903--3918},
	publisher = {American Physical Society},
	title = {Three-level quantum amplifier as a heat engine: A study in finite-time thermodynamics},
	url = {https://link.aps.org/doi/10.1103/PhysRevE.49.3903},
	volume = {49},
	year = {1994}}

@misc{Nayak2026,
    author = {Nayak, Pranay and Manikandan, Sreenath K. and Vu, Tan Van and Krishnamurthy, Supriya},
    title = {Signatures of coherent initial ensembles on all work moments},
    year = {2026},
    eprint = {2602.00227},
    archivePrefix = {arXiv},
    primaryClass = {quant-ph},
    doi = {10.48550/arXiv.2602.00227},
    url = {https://arxiv.org/abs/2602.00227}
}

@article{Manikandan2023,
  title = {Autonomous quantum clocks using athermal resources},
  author = {Manikandan, Sreenath K.},
  journal = {Phys. Rev. Res.},
  volume = {5},
  issue = {4},
  pages = {043013},
  numpages = {11},
  year = {2023},
  month = {Oct},
  publisher = {American Physical Society},
  doi = {10.1103/PhysRevResearch.5.043013},
  url = {https://link.aps.org/doi/10.1103/PhysRevResearch.5.043013}
}

@article{Fedorov2023,
  title = {Effect of Measurement Backaction on Quantum Clock Precision Studied with a Superconducting Circuit},
  author = {He, Xin and Pakkiam, Prasanna and Gangat, Adil A. and Kewming, Michael J. and Milburn, Gerard J. and Fedorov, Arkady},
  journal = {Phys. Rev. Appl.},
  volume = {20},
  issue = {3},
  pages = {034038},
  numpages = {17},
  year = {2023},
  month = {Sep},
  publisher = {American Physical Society},
  doi = {10.1103/PhysRevApplied.20.034038},
  url = {https://link.aps.org/doi/10.1103/PhysRevApplied.20.034038}
}

@article{DorfmanPark,
  title = {Coherence-enhanced quantum-dot heat engine},
  author = {Um, Jaegon and Dorfman, Konstantin E. and Park, Hyunggyu},
  journal = {Phys. Rev. Res.},
  volume = {4},
  issue = {3},
  pages = {L032034},
  numpages = {6},
  year = {2022},
  month = {Aug},
  publisher = {American Physical Society},
  doi = {10.1103/PhysRevResearch.4.L032034},
  url = {https://link.aps.org/doi/10.1103/PhysRevResearch.4.L032034}
}

@article{Tscherbul2014,
  title = {Long-Lived Quasistationary Coherences in a $V$-type System Driven by Incoherent Light},
  author = {Tscherbul, Timur V. and Brumer, Paul},
  journal = {Phys. Rev. Lett.},
  volume = {113},
  issue = {11},
  pages = {113601},
  numpages = {5},
  year = {2014},
  month = {Sep},
  publisher = {American Physical Society},
  doi = {10.1103/PhysRevLett.113.113601},
  url = {https://link.aps.org/doi/10.1103/PhysRevLett.113.113601}
}

@article{Dorfman2018,
	author = {Dorfman, Konstantin E. and Xu, Dazhi and Cao, Jianshu},
	doi = {10.1103/PhysRevE.97.042120},
	issue = {4},
	journal = {Phys. Rev. E},
	month = {Apr},
	numpages = {8},
	pages = {042120},
	publisher = {American Physical Society},
	title = {Efficiency at maximum power of a laser quantum heat engine enhanced by noise-induced coherence},
	url = {https://link.aps.org/doi/10.1103/PhysRevE.97.042120},
	volume = {97},
	year = {2018}}

@book{Mahler,
	author = {Mahler, G{\"u}nter},
	publisher = {Jenny Stanford Publishing},
	title = {Quantum thermodynamic processes: Energy and information flow at the nanoscale},
	year = {2014}}

@book{DeffnerBook,
	author = {Deffner, Sebastian and Campbell, Steve},
	publisher = {Morgan \& Claypool Publishers},
	title = {Quantum Thermodynamics},
	year = {2019}}

@incollection{AlickiKosloff,
	author = {Alicki, Robert and Kosloff, Ronnie},
	booktitle = {Thermodynamics in the Quantum Regime},
	pages = {1-33},
	publisher = {Springer},
	title = {Introduction to quantum thermodynamics: History and prospects},
	year = {2018}}

@article{Sourav2021,
	author = {Bhattacharjee, Sourav and Dutta, Amit},
	date = {2021/12/08},
	doi = {10.1140/epjb/s10051-021-00235-3},
	id = {Bhattacharjee2021},
	isbn = {1434-6036},
	journal = {The European Physical Journal B},
	number = {12},
	pages = {239},
	title = {Quantum thermal machines and batteries},
	url = {https://doi.org/10.1140/epjb/s10051-021-00235-3},
	volume = {94},
	year = {2021}}

@article{Sai2016,
	author = {Vinjanampathy, S. and Anders, J.},
	doi = {10.1080/00107514.2016.1201896},
	journal = {Contemporary Physics},
	number = {4},
	pages = {545-579},
	publisher = {Taylor & Francis},
	title = {Quantum thermodynamics},
	volume = {57},
	year = {2016}}

@article{Terlizzi,
    author = {Di Terlizzi, Ivan and Baiesi, Marco},
    journal = {J. Phys. A: Math. Theor.},
    volume = {52},
    number = {2},
    pages = {02LT03},
    title = {Kinetic uncertainty relation},
    year = {2019},
    doi = {10.1088/1751-8121/aaee34},
    url = {https://doi.org/10.1088/1751-8121/aaee34}
}

@article{SinghKwonLeeKUR2026,
  author = {Singh, Varinder and Kwon, Euijoon and Lee, Jae Sung},
  title = {Violation of kinetic uncertainty relation in maser heat engines: Role of spontaneous emission},
  journal = {Phys. Rev. A},
  volume = {113},
  pages = {012212},
  year = {2026},
  doi = {10.1103/qgpx-3pkf},
  url = {https://doi.org/10.1103/qgpx-3pkf}
}

@article{Dechant_2018,
doi = {10.1088/1742-5468/aac91a},
url = {https://doi.org/10.1088/1742-5468/aac91a},
year = {2018},
month = {jun},
publisher = {IOP Publishing and SISSA},
volume = {2018},
number = {6},
pages = {063209},
author = {Dechant, Andreas and Sasa, Shin-ichi},
title = {Current fluctuations and transport efficiency for general Langevin systems},
journal = {Journal of Statistical Mechanics: Theory and Experiment}
}

@article{Hasegawa2019_underdamped,
  title = {Uncertainty relations for underdamped Langevin dynamics},
  author = {Van Vu, Tan and Hasegawa, Yoshihiko},
  journal = {Phys. Rev. E},
  volume = {100},
  issue = {3},
  pages = {032130},
  numpages = {13},
  year = {2019},
  month = {Sep},
  publisher = {American Physical Society},
  doi = {10.1103/PhysRevE.100.032130},
  url = {https://link.aps.org/doi/10.1103/PhysRevE.100.032130}
}

@article{JSLee2021_universalTUR,
  title = {Universal form of thermodynamic uncertainty relation for Langevin dynamics},
  author = {Lee, Jae Sung and Park, Jong-Min and Park, Hyunggyu},
  journal = {Phys. Rev. E},
  volume = {104},
  issue = {5},
  pages = {L052102},
  numpages = {5},
  year = {2021},
  month = {Nov},
  publisher = {American Physical Society},
  doi = {10.1103/PhysRevE.104.L052102},
  url = {https://link.aps.org/doi/10.1103/PhysRevE.104.L052102}
}

@article{Kwon2024_unified,
  title = {Unified hierarchical relationship between thermodynamic tradeoff relations},
  author = {Kwon, Euijoon and Park, Jong-Min and Lee, Jae Sung and Baek, Yongjoo},
  journal = {Phys. Rev. E},
  volume = {110},
  issue = {4},
  pages = {044131},
  numpages = {20},
  year = {2024},
  month = {Oct},
  publisher = {American Physical Society},
  doi = {10.1103/PhysRevE.110.044131},
  url = {https://link.aps.org/doi/10.1103/PhysRevE.110.044131}
}

@article{Brandner2025,
  title = {Thermodynamic Uncertainty Relations for Coherent Transport},
  author = {Brandner, Kay and Saito, Keiji},
  journal = {Phys. Rev. Lett.},
  volume = {135},
  issue = {4},
  pages = {046302},
  numpages = {7},
  year = {2025},
  month = {Jul},
  publisher = {American Physical Society},
  doi = {10.1103/6nww-8wcp},
  url = {https://link.aps.org/doi/10.1103/6nww-8wcp}
}

\bibliographystyle{apsrev4-2}

\end{document}